\documentclass[12pt]{article}
\usepackage{psfig}
\usepackage{graphics}

\newcommand{\be}{\begin{equation}}
\newcommand{\ee}{\end{equation}}
\newcommand{\bea}{\begin{eqnarray}}
\newcommand{\eea}{\end{eqnarray}}

\newcommand{\tl}{\tilde{\ell}}
\newcommand{\cl}{\ell}
\newcommand{\cQ}{{\mathcal{Q}}}

\newcommand{\cB}{{\mathcal{B}}}
\newcommand{\cE}{{\mathcal{E}}}

\newcommand{\cF}{{\mathcal{F}}}

\newcommand{\tN}{{\tilde{N}}}

\newcommand{\cH}{{\mathcal{H}}}

\newcommand{\cR}{{\mathcal{R}}}

\newcommand{\cA}{{\mathcal{A}}}  

\newcommand{\cM}{{\mathcal{M}}}

\newcommand{\ssg}{{\sqrt{\sigma}}}

\newcommand{\nn}{\nonumber}

\newcommand{\az}{{\alpha}}
\newcommand{\bz}{{\beta}}
\newcommand{\cz}{{\gamma}}

\newcommand{\dz}{{\delta}}

\newcommand{\bu}{{\bar{u}}}
\newcommand{\bn}{{\bar{n}}}
\newcommand{\bN}{{\bar{N}}}
\newcommand{\bV}{{\bar{V}}}
\newcommand{\bk}{{\bar{k}}}
\newcommand{\ba}{{\bar{a}}}
\newcommand{\bep}{{\bar{\varepsilon}}}
\newcommand{\bj}{{\bar{\jmath}}}
\newcommand{\bs}{{\bar{s}}}

\newcommand{\bPhi}{{\bar{\Phi}}}

\newtheorem{defn}{Definition}

\begin{document}

\title{Metric-based Hamiltonians, null boundaries, and isolated horizons}
       
\author{Ivan S. Booth\footnote{ibooth@phys.ualberta.ca} \\
                               Theoretical Physics Institute\\
                               Department of Physics\\
                               University of Alberta\\
                               Edmonton, Alberta T6G 2J1\\
                               Canada}
\date{May 8, 2001\\}
\maketitle
         
\begin{abstract}
We extend the quasilocal (metric-based) Hamiltonian formulation of 
general relativity so that it may be used to study regions of spacetime 
with null boundaries. In particular we use this generalized Brown-York 
formalism to study the physics of isolated horizons. We show that 
the first law of isolated horizon mechanics follows directly from the
first variation of the Hamiltonian. This variation is not restricted to
the phase space of solutions to the equations of motion but is instead
through the space of all (off-shell) spacetimes that contain isolated 
horizons. We find two-surface integrals evaluated on the horizons that are
consistent with the Hamiltonian and which define the energy and angular 
momentum of these objects. These are closely related to the corresponding 
Komar integrals and for Kerr-Newman spacetime are equal to the corresponding 
ADM/Bondi quantities. Thus, the energy of an isolated horizon calculated
by this method is in agreement with that recently calculated by Ashtekar and 
collaborators but not the same as the corresponding quasilocal energy defined 
by Brown and York. Isolated horizon mechanics and Brown-York thermodynamics 
are compared.
\end{abstract}

\section{Introduction}
In the almost 30 years since its initial discovery by Bekenstein \cite{bek}
and Hawking \cite{hawk1}, black hole thermodynamics has been one of the most
active areas of gravitational research. There have been several
reasons for this interest, but perhaps the chief one is that these laws
represent one of the few solid links between the classical world of general
relativity and the much-sought-after theory of quantum gravity. 
Just as quantum mechanics gives rise to the
statistical mechanics that in turn explains the  
thermodynamics of condensed matter, it is believed that the 
laws of quantum gravity will explain the statistical mechanics and therefore
thermodynamics of black holes. The advances made by string theory and
canonical quantum gravity towards this goal in the last few years are
well known (see \cite{quantG} for a review and references).

That said, there are several ways in which standard black hole thermodynamics
differs from the more familiar thermodynamics of everyday materials. Perhaps
the most obvious problem with black holes comes from their very definition.
While one can easily localize standard thermodynamic systems such as
steam engines in spacetime, this is not the case for a black hole.
By the standard definition, a black hole is a region of spacetime 
from which no null or timelike signal can ever escape. Thus, one must know 
the entire history of a spacetime before one can say with certainty whether
or not a particular region is inside a black hole. Such a definition is too
unwieldy to use except in the very special case of stationary spacetimes and so
one would like to find a more general class of objects that retain the 
properities of black holes that are essential for thermodynamics, while at the
same time being more convenient to identify and study.

A second problem with black hole thermodynamics arises in the definition
of thermodynamic quantities such as energy or angular momentum. While
the entropy of a black hole is measured at the event horizon itself,
in the usual formulation of black hole thermodynamics energy and angular 
momentum are defined by ADM measurements taken back at infinity. Further,
to define the temperature of a black hole one must make reference to Killing 
vector fields at infinity. 
Again this is a situation that is alien to less esoteric thermodynamics. 
One can define the energy of a standard thermodynamic system such as 
an ice-cube without 
first having to retreat to the outer reaches of the universe. A local
definition of energy is obviously important too if one doesn't want to
include stray galaxies sitting megaparsecs away from a black hole in
its total mass. 
Thus, to give a good formulation of gravitational thermodynamics, 
we need quasilocal ways to both identify black holes (or some generalization
of them) and to measure all of their thermodynamic properties.

Quite recently Ashtekar, Beetle, and Fairhurst proposed a solution to these
problems when they defined isolated horizons \cite{ABF1,ABF2}.
Roughly these are closed, non-expanding, null surfaces which are in 
equilibrium with their surrounding spacetime (though as long as the rest of
the spacetime doesn't disturb the equilibrium of the horizon it may itself
be dynamic). In a series of papers following
the initial two, those authors and others have studied the geometric
properties of various species of isolated horizons and also shown how they
naturally fit into the tetrad and spinor Hamiltonian formulations of general
relativity. They have provided quasilocal, horizon-based definitions of the 
thermodynamic properties of isolated horizons and used the Hamiltonian 
formalism to derive the zeroth and first laws of thermodynamics
(see for example \cite{IHlett} for a review of this work and further 
references). 

Of course, Hamiltonian derived definitions of quasilocal energy and even 
quasilocal thermodynamics are nothing new and there has been an 
active literature on these subjects for some years (see for example
\cite{BY1, QLE} and the references contained therein). 
In fact, the quasilocal energy community has generally been more
ambitious than the isolated horizon one and sought to define the energy
of arbitrary regions of spacetime rather than just black holes. Unfortunately,
the various definitions tend to disagree on their localizations and no 
consensus on what is the ``correct'' quasilocal energy (or even if such a
thing exists) has been reached. Nevertheless, there are well developed
quasilocal formalisms out there, and it is of interest to see what they 
have to say about isolated horizons and how their quasilocal 
definitions relate to the isolated horizon ones.

One of the most popular definitions of quasilocal energy \cite{BY1}
and formulations of quasilocal thermodynamics \cite{BY2} was proposed 
by Brown and York in 1993. Starting with the standard 
Einstein-Hilbert action (with boundary terms) they calculated the first
variation of the action. Setting it equal to zero they obtained the equations
of motion (in the usual way) plus a set of boundary conditions. Enforcing 
these boundary conditions causes the variational boundary term to vanish and
so ensure that the variational principle is well-defined. 
The conditions vary according to the
original boundary terms of the action and so there is quite a bit of freedom
in the whole procedure. That said, for
each choice of the action, one can apply a Legendre transform, obtain
a Hamiltonian defined on spatial three-surfaces, and so define the 
quasilocal energy (QLE) of any finite section of that hypersurface 
as the numerical value of the Hamiltonian evaluated over the region. The
trick comes in picking the correct action and it turns out that even in
the most simple cases, there is some ambiguity in the definition of the QLE.
This ambiguity is interpreted as the freedom to pick the zero-point of the 
energy. Similar comments may be made about angular momentum, though there
the zero-point ambiguity doesn't arise. 

This formalism has many nice characteristics. For example, it reduces to the 
ADM \cite{BY1} and Bondi \cite{BYBondi} energies in the appropriate
infinite limits, while in the limit of a sufficiently small quasilocal region
with volume $V$, it reduces to the matter energy 
$ (V \times T_{\az \bz} u^\az u^\bz)$ where $T_{\az \bz}$ and 
$u^\az$ are respectively local values of the stress-energy tensor and the
timelike normal to the spacelike hypersurface \cite{BYsmall}. With the help of
Euclidean path integral gravity one can use it to (partially) 
explain black hole thermodynamics (appendix \ref{A} or \cite{BY2} for more 
details). 
It has even been used to 
study gravitational tidal heating \cite{tidal}.
However, the formalism as it stands is not quite suitable for studying
isolated horizons. Specifically, quasilocal quantities such as
energy and angular momentum are defined by sets of observers who evolve
in a timelike way. For an isolated horizon, we want to make measurements  
on the null surface itself which means that such observers must
have null (or even spacelike) velocities\footnote{The reader
might object that physical observers cannot travel with such velocities. 
That is a valid point. Nevertheless from an intuitive point of view it is
convenient to continue to think of the boundaries as being made up of
observers.}.

In this paper, we extend the metric-based Hamiltonian perspective to 
include null boundary surfaces (and classes of observers moving with
null or spacelike velocities on those surfaces). The extension is fairly
straightforward from a geometrical point of view, but nevertheless
the final results differ quite dramatically from those for timelike
boundaries. In the standard Brown-York work quasilocal energy (sometimes
referred to as the canonical quasilocal energy or CQLE) 
is essentially measured by the extrinsic curvature of two-surfaces 
surrounding the system. By contrast, in the new null formulation the QLE 
of an  isolated horizon (which we'll denote the null QLE or NQLE)
is measured not by the expansion of the
congruence of null curves that make up the surface 
(which would be the naive analogue of the extrinsic curvature) 
but rather by the acceleration of those same curves. Further,
while the CQLE suffers from the ambiguity in the aforementioned
zero-point terms, for the NQLE this freedom of choice is largely removed by 
Hamiltonian formalism. This is in agreement with the isolated horizon
literature (see for example \cite{ABF2}). 
Finally, the NQLE measured on an isolated horizon is 
NOT equal to the CQLE measured on a timelike surface positioned
``close'' to the horizon. Specifically, in the case of Schwarzschild,
the NQLE is $m$ (in agreement with \cite{ABF1}) while the 
CQLE is $2m$. In fact, for any stationary, axisymmetric isolated horizon
in asymptotically flat space, the NQLE is equal to the ADM/Bondi mass of the
entire spacetime.

In other ways, however, the results correspond much more closely to those
of Brown and York. In particular, the expression for angular momentum on 
a timelike surface is very closely related to the corresponding expression
on a null surface (which in turn we show to be eqivalent to the corresponding
Komar formula). Further, basically the same Hamiltonian boundary terms arise
on an isolated horizon as those that show up on a timelike surface, though as 
noted above the
roles that they play can be quite different. Also, drawing on 
the extant work on metric-based Hamiltonians, we can, without much difficulty,
carry through the variational calculations explicitly for the phase space
of all spacetime configurations rather than just the 
subspace of solutions to the equations of motion. 

This paper is organized in the following manner. Section \ref{setup}
establishes notation and sign conventions, reviews the definitions and 
some properties of isolated horizons and considers the 
constants that characterize these objects. The notation will
be an amalgam of that used in the Brown-York tradition of papers with that
used in studies of isolated horizons with some new symbols introduced where
the systems overlap and clash. Section \ref{calcs} extends the
the work on metric-based Hamiltonians to include systems with null boundaries
in general and rigidly rotating horizons (a close cousin of isolated horizons)
in particular. We find constraints on possible boundary terms for the 
Hamiltonian. Section \ref{phys}, derives the first law of isolated
horizon mechanics, proposes an integral boundary term for the Hamiltonian
that both satisfies the constraints and is equivalent to the function found in 
\cite{ABL}, and considers the connection 
between the quasilocal energy/angular momentum and the corresponding Komar
definitions. Section \ref{BY} compares the results found for
isolated horizons in this paper with the Brown-York quasilocal energy and 
thermodynamics. Section 
\ref{discuss} summarizes the results and appendix \ref{A} reviews
Brown-York thermodynamics.

\section{Set-up}
\label{setup}

The calculations in this paper will be done from a Hamiltonian perspective.
As such it will usually be most appropriate to think not of spacetime,
but instead of time-dependent tensor fields living and evolving on a 
spatial three-manifold. At the same time however, many of the quantities
appearing in the calculations can most easily be understood as 
four-dimensional quantitities associated with three-surfaces embedded in 
a full spacetime. As such, and even though it will create a certain amount
of clutter, we will begin by setting things up from a three-dimensional 
perspective and then ``stack'' the ``instants'' of time evolution into a 
full four-dimensional spacetime. Only then will we define the various
sub-species of isolated horizons and then proceed on to the variational
calculations. Note that throughout this paper, we'll use units where
$G$ and $c$ are dimensionless and equal to unity.

\subsection{The 3D perspective}
\begin{figure}
\centerline{\psfig{figure=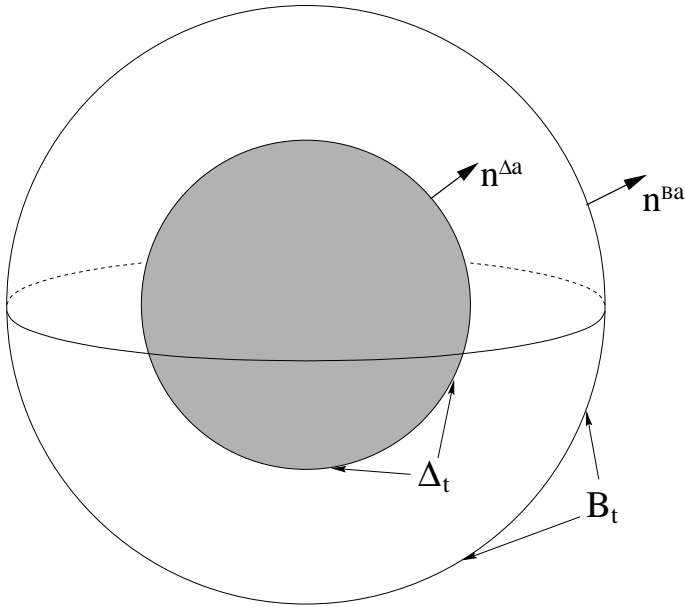,height=8cm,angle=270}}
\caption{The region $\Sigma_t$ with its boundaries and normals.}
\label{fig1}
\end{figure}

We begin with the three-dimensional-space plus time perspective, 
viewing fields as defined in a three-dimensional manifold
$\Sigma$ and dynamic with respect to a time coordinate $t$. To
emphasize their three-dimensional nature we label them with Latin
indices. Figure \ref{fig1} illustrates this section.

Let $\Sigma$ be a three-dimensional manifold with Euclidean metric
$h_{ab}$, and $\Sigma_t \subset \Sigma$ be
such that $\Sigma_t$ is topologically $S^2 \times [0,1]$ (where 
$S^2$ is the two-sphere and $[0,1]$ is a real line segment). 
Label the 
two-dimensional boundaries of this space $\Delta_t$ and 
$B_t$ and refer to them as the ``inner'' and ``outer'' boundaries. 
The induced metrics on these boundaries will be
$\sigma^\Delta_{ab}$ and $\sigma^B_{ab}$ respectively. Next, if $D_a$ is the
metric compatible covariant derivative on $\Sigma$ then the corresponding
induced metrics on the boundaries will be $d_a^\Delta$ and $d_a^B$ 
respectively. Often where there is no chance of confusion we will drop the 
superscripts on these quantities.

Next, define $n^\Delta_a$ and $n^B_a$ as the unit normals to the 
two-surfaces. They are oriented so that $n^\Delta_a$ points back ``into'' 
and $n^B_a$ points ``out of'' $\Sigma_t$. Then, one can define
the extrinsic curvatures 
$k^\Delta_{ab} = - \frac{1}{2} \mathcal{L}_{n} \sigma^\Delta_{ab} 
= - \sigma^{\Delta c}_a \sigma^{\Delta d}_b D_c n^\Delta_d$ 
and $k^B_{ab} = - \frac{1}{2} \mathcal{L}_{n} \sigma^B_{ab} 
= - \sigma^{B c}_a \sigma^{B d}_b D_c n^B_d$, 
where $\mathcal{L}_v$ is used to designate a Lie 
derivative in the direction $v$. The reader should take note of
the sign convention used for extrinsic curvatures. 

The lapse function $N$ and shift vector $V^a$ will define the flow
of time relative to this surface in the usual way. That is,
$N dt$ units of proper time pass for every $dt$ units of coordinate time, and 
an observer swept along by the flow of time will move with velocity 
$\sqrt{V_a V^a}/N$ 
\footnote{Keep in mind that depending on the spacetime, such ``motion'' may
or may not correspond to ``real'' movement. For example, a particle which 
is static on a Schwarzschild horizon will have a velocity 
equal to the speed of light, but doesn't actually move by any normal 
definition of movement.}.
In particular we will think of the boundaries
$\Delta_t$ and $B_t$ as being made up of such observers, 
and so their evolution will be guided by the shift vector field.  
Thus, given the position of $\Delta_t$ and $B_t$ at any instant $t$, the
flow of time determines their position in $\Sigma$ 
at all future times and therefore the future extent of $\Sigma_t$ as well. 

In addition to the metric field $h_{ab}$, the Hamiltonian three-surface
formulation of general relativity requires a surface-momentum tensor
density $P^{ab}$ which is canonically conjugate to the metric (and  
determines the time rate of charge of the metric). Further, for
Einstein-Maxwell spacetimes, there will also be the
vector potential one-form $A_a$, its canonical conjugate 
the electric field vector density $\cE^a$ (which apart from its
obvious interpretation measures the time rate of change of $A_a$),
and the Coulomb potential
$\Phi$ (which joins the lapse and shift as a Lagrange multiplier). 
 
We now embed all of this in a four-dimensional spacetime.

\subsection{The 4D perspective}
\label{SetUp4D}

\begin{figure}
\centerline{\psfig{figure=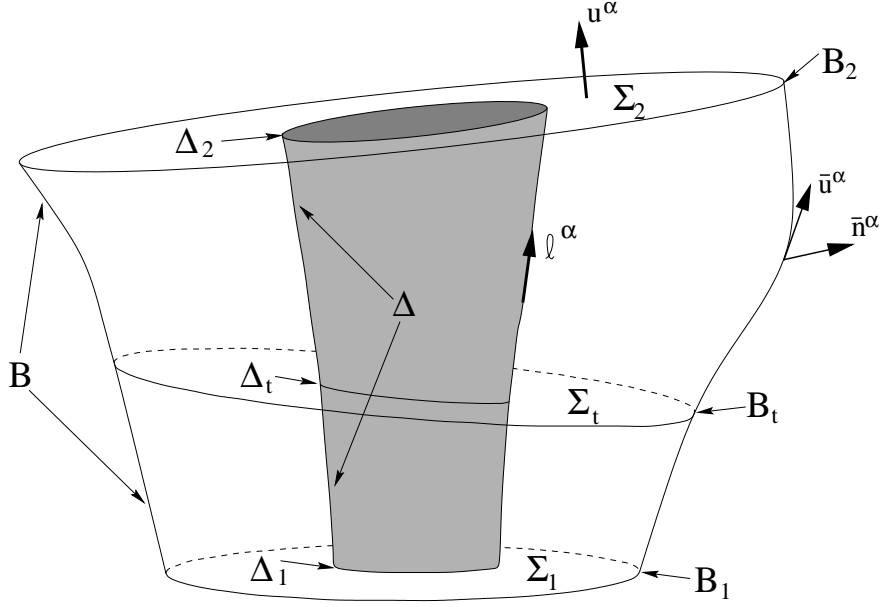,height=8cm,angle=270}}
\caption{A three-dimensional schematic of $M$ along with its boundaries,
normals, and a typical foliation surface. 
No attempt has been made to have the slope of the boundaries
correspond to their ``speed''.}
\label{fig2}
\end{figure}

Let $\cM$ be a four-dimensional manifold and $M \subset \cM$ such that
$M$ is topologically $[0,1] \times [0,1] \times S^2$. 
Thus, $M$ is ``cylindrical''
and bounded by four three-dimensional surfaces $\Sigma_1$,
$\Sigma_2$, $\Delta$, $B$ (see figure \ref{fig2}). Next, let $g_{\az \bz}$ be
the metric field on $\cM$ such that $\Sigma_1$ and $\Sigma_2$ are spacelike, 
$\Delta$ is null, and $B$ is timelike. Note that we use Greek indices for 
these four-dimensional quantities. Other situations will be
considered later on in the text, but this set-up will allow us to 
capture all of the essential details for now and for later.

Assume that $M$ is foliated by a set of spacelike surfaces 
$\{\Sigma_t : t_1 \leq t \leq t_2\}$ such that 
$\Sigma_{t_1} = \Sigma_1$ and $\Sigma_{t_2} = \Sigma_2$ are the 
terminal elements of the foliation. These surfaces can then be
thought of as ``instants'' in time and in the usual way, we'll say
that $\Sigma_{t_\az}$ ``happens'' before $\Sigma_{t_\bz}$ if 
$t_\az < t_\bz$. A further orientation is implied by choosing to refer
to $\Delta$ as the inner boundary and  $B$ as the outer boundary. 
The $\Sigma_t$ then induce corresponding foliations
$\{ \Delta_t : t_1 \leq t_2 \} $ and 
$\{ B_t : t_1 \leq t \leq t_2 \}$ on $\Delta$ and $B$ respectively. 
In the usual ADM/Brown-York formulation of general relativity, these foliations
would be independent of the spacetime configuration. However, that will not
be the case here and in section \ref{calcs} we will see how the Hamiltonian
formalism restricts the allowed foliations of $\Delta$ (and therefore
$M$) if $\Delta$ is an isolated horizon.

Using the spacetime metric $g_{\az \bz}$ one may define normal vectors, 
projection operators, 
and metric tensors for the hypersurfaces. Specifically, if $u^\az$
is the forward pointing unit timelike normal to the $\Sigma_t$ then 
$h_\az^\bz = g_\az^\bz + u^{\az} u_\bz$ is the projection operator
into the surface, while $h_{\az \bz} = g_{\az \bz} + u_\az u_\bz$ is the
induced spacelike metric. 
Similarly, if $\bn^\az$ is the outward pointing unit normal to 
$B$ then $\gamma_\az^\bz = g_\az^\bz - \bn_\az \bn^\bz$ is the
projection operator, and 
$\gamma_{\az \bz} = g_{\az \bz} - \bn_\az \bn_\bz$ is the 
induced timelike metric.
For the two-surfaces $B_t$,
the projection operator maybe written as either of
$\sigma_\az^\bz = h_\az^\bz - n_\az n^\bz = \gamma^\az_\bz + \bu^\az \bu_\bz$
where $n_\az$ is the outward pointing spacelike unit normal to 
$B_t$ in $\Sigma_t$ and
$\bu_\az$ is the future pointing timelike unit normal to $B_t$ in $B$.
The induced metric is then 
$\sigma_{\az \bz} =  h_{\az \bz} - n_\az n_\bz = \gamma_{\az \bz} 
+\bu_\az \bu_\bz$. Note that $\bn_\az \neq n_\az$ and $\bu_\az \neq u_\az$
unless the foliation surfaces are perpendicular to $B$. That is 
$\eta \equiv \bu^\az n_\az = 0$\footnote{Note that the sign convention for 
$\eta$ agrees with \cite{mythesis} and the transformation law section
of \cite{naked} but is the opposite from that originally used by 
Hayward \cite{hayward} and the original non-orthogonal boundary paper
cowritten by the author \cite{nopaper}.}.

Next, let $\cl_\az$ be a null normal to $\Delta$. Thus if $\Delta$ is
defined as a level surface of some function $H$ of the spacetime coordinates,
$\cl_\az$ is proportional to $\partial_\az H$. Further, assume that
it points foward in time and into $M$ so that 
$\ell^\az u_\az <0$ and $\ell^\az n_\az > 0$. However, because
$\ell_\az$ is null, it has no natural normalization.
To define the projection operator into $\Delta$, one must invoke an auxilarly 
null normal $\tl^\az$ which is also assumed to be perpendicular to $\Delta_t$ 
and which is normalized relative to $\cl_\az$ so that 
$\tl^\az \cl_\az = -1$. Then if  $n_\az$ is the inward-pointing spacelike 
unit normal to $\Delta_t$, one can write 
\be
\ell_\az = \xi (u_\az + n_\az) \hspace{1cm} \mbox{and} \hspace{1cm}
\tl_\az = \frac{1}{2 \xi} (u^\az - n^\az)
\ee
for some function $\xi$. 
That said, the projection operator
into $\Delta$ is $q_\az^\bz = g_\az^\bz + \tl_\az \cl^\bz$ and the 
induced metric is $q_{\az \bz} = g_{\az \bz} + \cl_\az \tl_\bz
+ \tl_\az \tl_\bz$. This metric is, of course degenerate and therefore
$q_{\az \bz}$ does not uniquely define its inverse $q^{\az \bz}$ (though this
non-uniqueness doesn't manifest itself in any of the quantities that 
we use in this paper). It should
also be kept in mind that $\cl^\az$ and $\tl_\az$ lie in the tangent
and cotangent bundles to $\Delta$ respectively while their 
compatriots $\tl^\az$ and $\cl_\az$ do not. The projection operator into 
$\Delta_t$ is
$\sigma_\az^\bz = g_\az^\bz + \cl_\az \tl^\bz + \tl_\az \cl^\bz
= g_\az^\bz + u_\az u^\bz - n_\az n^\bz$. The induced metric is then 
$\sigma_{\az\bz} = g_{\az \bz} + \cl_\az \tl_\bz + \tl_\az \cl_\bz
= g_{\az \bz} + u_\az u_\bz - n_\az n_\bz$. Note that 
$\sigma_{\az \bz} = q_{\az \bz}$.

The metric compatible covariant derivative on $\cM$ will be $\nabla_\az$ 
which in turn will induce $D_\az$ as the metric compatible derivative on 
$\Sigma_t$, $d^\Delta_\az$ on $\Delta_t$, and $d^B_\az$ on $B_t$.
$\triangle_\az$ will be the induced covariant derivative on $\Delta$. 
Note that it is compatible with $q_{\az \bz}$, 
though not uniquely determined by it since that metric is degenerate.

The extrinsic curvatures of these hypersurfaces may be defined as follows. 
$K_{\az \bz} = - \frac{1}{2} \mathcal{L}_u h_{\az \bz} = 
- h_\az^\cz \nabla_\bz u_\cz$ is the
extrinsic curvature of $\Sigma_t$ in $M$, 
$k_{\az \bz} = - \sigma_\az^\cz \sigma_\bz^\dz \nabla_\cz n_\dz$, is the
extrinsic curvature of $B_t$ in $\Sigma_t$, and 
$l_{\az \bz} =  - \sigma_\az^\cz \sigma_\bz^\dz \nabla_\cz u_\dz$ is the
extrinsic curvature of $B_t$ with respect to the normal $u_\az$. Each 
of these last two quantities may be considered to be defined for $\Delta_t$ 
or $B_t$ as appropriate. Contracted with the appropriate metric tensors
these become $K$, $k$, and $l$. The addition of a bar over any of these
indicates that they are to be calculated with respected to the ``barred'' 
rather than ``unbarred'' normal. For example $\bar{l} = - \sigma^{\az \bz} 
\nabla_\az \bu_\bz$. 

On $\Delta$, the analogous quantity is  
$\theta_{\az \bz} =  q_\az^\cz q_\bz^\dz \nabla_\cz \ell_\dz
= \sigma_\az^\cz \sigma_\bz^\dz \nabla_\cz \ell_\dz$,
which describes the instaneous evolution of the congruences of null 
geodesics that make up $\Delta$. The reader should take note of the sign
convention which is different from that used in the extrinsic curvatures.
Then, keeping in mind that $\ell_\az$ is normal to $\Delta$ so that 
$\theta_{[\alpha \beta]}=0$, this may be decomposed as 
\be
\label{thetaab}
\theta_{\az \bz} = \frac{1}{2} \theta \sigma_{\az \bz} + \zeta_{\az \bz}
\ee
where $\theta \equiv \sigma^{\az \bz} \theta_{\az \bz} 
= \frac{1}{\ssg} \mathcal{L}_\ell \ssg$ is the null expansion and 
$\zeta_{\az \bz} \equiv \theta_{\az \bz} - \frac{1}{2} \theta \sigma_{\az \bz}$
is the shear. If $\theta_{\az \bz} = 0$, then $q_{\az \bz}$ 
has a unique compatible covariant derivative $\triangle_\az$ on $\Delta$. 

The flow of time in this manifold is defined by the vector field
$T^\az$ which is chosen to be compatible with the foliation $\{\Sigma_t\}$ so 
that $T^\az \partial_\az t = 1$. On the hypersurfaces it may be written 
as $T^\az = N u^\az + V^\az$ where again $N$ and $V^\az$ are the lapse and the
shift. It is not restricted to being timelike. In fact, near or on $\Delta$ 
we will
expect it to go null or even spacelike. Because of the restriction made in the
previous section that the evolution of the boundaries by guided by the time
evolution, we know that $T^\az$ may be written as an element of the
tangent bundles $T\Delta$ and $TB$ at the inner and outer 
boundaries respectively. Specifically, on $\Delta$ we have
$T^\az = \tilde{N} \ell^\az + \hat{V}^\az$ where 
$\tilde{N} \equiv N/\xi$ and $\hat{V}^\az \equiv \sigma^\az_\bz V^\bz$, while
on $B$, $T^\az = \bar{N} \bu^\az + \hat{V}^\az$ where 
$\bar{N} \equiv N \lambda$, $\lambda \equiv \sqrt{1+ \eta^2}$, and (again)
$\hat{V}^\az \equiv \sigma^\az_\bz V^\bz$. In section \ref{calcs} we will see
that the allowed forms of $T^\az$ are restricted at isolated horizons. Again 
this contrasts with the usual ADM/Brown-York Hamiltonian formulations where 
$T^\az$ is an independent background structure.

From this perspective the ``instants'' are a series of hypersurfaces,
rather than a single three-manifold on which dynamic fields evolve. It
will be useful to note that the hypersurface momentum $P^{ab}$ which was 
an independent variable in the previous section, is closely related to the
extrinsic curvature $K_{\az \bz}$ and can be written as
\be
P^{\az \bz} = \frac{\sqrt{h}}{16  \pi} (K h^{\az \bz} - K^{\az \bz}).
\ee 
The electromagnetic field in $\cM$ is described by the vector potential
$\cA_\az$ which defines the electromagnetic field tensor 
$\cF_{\az \bz} = \partial_\az \cA_\bz - \partial_\bz \cA_\az$. Then, the
three-dimensional vector potential is $A_\az = h_\az^\bz \cA_\bz$, the
Coulomb potential is $\Phi = - \cA_\az u^\az$, and the
electric field density is 
$\cE^\az \equiv - \sqrt{h}/(4 \pi ) \cF^{\az \bz} u_\bz$. Throughout this
paper we will assume that a single $\cA_\az$ can be used to cover the entire 
region $\cM$. This assumption means that the electromagnetic $U(1)$ gauge
bundle will be trivial and therefore there are no magnetic charges 
in $\cM$ -- including magnetically charged black holes. Results for 
magnetically charged spacetimes can be found using electromagnetic duality.

It should also be kept in mind that constant $t$ surfaces in Scwharzschild/Kerr
spacetime in the usual static/stationary coordinates are NOT examples of 
the $\Sigma_t$ surfaces that we are considering here. 
As is well known, such constant $t$ surfaces go null 
at the event horizon and timelike inside of it. Our 
$\Sigma_t$ are restricted to remain spacelike and so one should instead
think of a foliation such as that induced by Painleve-Gullstand-type
coordinates. These are analogous to Eddington-Finkelstein coordinates
except that they are tied to infalling timelike rather than
spacelike geodesics. Descriptions of these in 
spherically symmetric and Kerr spacetimes may be found in
\cite{PGeric} and \cite{PGkerr} respectively.

\subsection{Review of isolated horizons}
\label{IHreview}
This section reviews the definitions of the isolated horizon sub-species 
as well as some of their properties. It is largely
based on the relevant sections of \cite{AFK} and \cite{ABL} and the reader is
directed there for further discussion and derivation of the definitions and 
results.

\subsubsection{Definitions}
\label{defns}

\begin{defn}
A null surface $\Delta$ is a 
{\bf non-expanding horizon} if: i) its null expansion $\theta$ is
zero everywhere on $\Delta$ for any null normal $\ell^\az$ 
ii) all the equations of motion hold at $\Delta$, and 
iii) the stress tensor $T_{\az \bz}$ of matter fields
is such that $- T_\az^\bz \ell^\az$ is future directed and causal for any 
null normal $\ell^\az$. 
\end{defn}

A couple of notes are in order on these conditions. First, it is trivial to
check that if conditions i) and iii) hold for any null normal vector field 
$\ell^\az$ then they will hold for all such vector fields and so the definition
isn't based on the normalization of $\ell^\az$. Second, condition iii) is 
implied by the dominant energy condition which in particular holds for the
Maxwell fields that we consider in this paper. 
Thus, from our point of view, the only real 
restriction in the above is that $\theta = 0$. Note that any Killing 
horizon is automatically a non-expanding horizon.

From this simple assumption, many properities follow. First $\Delta$ may be 
thought of as a congruence of null geodesics. By definition they are 
surface forming and so their twist 
$\theta_{[\az \bz]}$ is zero. 
With the help of the Raychaudri equation one can quickly see that $\theta=0$ 
then implies that their shear $\zeta_{\az \bz}$ must be zero too. 
Therefore $\theta_{\az \bz} = 0$.
It then follows that the induced $\triangle_\az$ is the unique compatible
covariant derivative for $q_{\az \bz}$. 
From the same calculation the twice contracted stress-energy tensor
$T^{\az \bz} \ell_\az \ell_\bz$ = 0 as well. 
In turn, $T^{\az \bz} \ell_\az \ell_\bz = 0$
implies that $f_\az \equiv \sigma_\az^\cz F_{\cz \dz} \ell^\dz$ and
$\tilde{f}_\az \equiv \sigma_\az^\cz F_{\cz \dz} \tl^\dz$ are both zero
(or equivalently the electric and magnetic fields 
projected into $\Delta_t$ are zero). 
Physically this means that there is no flux of electromagnetic 
radiation across the horizon (the relevant component of the Poynting
vector is zero). 

Moving on, direct calculation and $\theta_{\az \bz} = 0$ implies
that there exists a one form $\omega_\az$ such 
that $q_\az^\cz \nabla_\cz \ell^\bz = \omega_\az \ell^\bz$. Explicitly,
\be
\omega_\az = - \kappa_\ell \tl_\az + \sigma_\az^\bz \ell^\cz \triangle_\bz
\tl_\cz \label{omega}
\ee
where $\kappa_\ell \equiv - \tl_\az \ell^\bz \triangle_\bz \ell^\az$ is the
acceleration of the null normal $\ell^\az$. Note that $\omega_\az$ is 
defined entirely with respect to quantities intrinsic to $\Delta$. 
These quantities are, however, dependent on the normalization of $\ell^\az$ and
under a redefinition of the null normal $\ell^\az \rightarrow \ell'^\az = 
f \ell^\az$, $\kappa_{\ell'} = f \kappa_\ell + \mathcal{L}_\ell f$ and
$\omega'_\az = \omega_\az + q_\az^\bz \nabla_\bz f$. 

\begin{defn}
A non-expanding horizon $\Delta$ becomes a 
{\bf weakly isolated horizon} 
$(\Delta,[\ell])$ if it is equipped with an 
equivalence class of normals $[\ell]$ such that 
$q_\az^\bz \mathcal{L}_\ell \omega_\bz = 0$ for all $\ell \in [\ell]$ (where
$\ell' \sim \ell$ if $\ell' = c\ell$ for some constant $c$) and a gauge choice
such that $q_\az^\bz \mathcal{L}_\ell \cA_\bz = 0$.  
\end{defn}

It quite easy to show that for any non-expanding horizon $\Delta$ 
there are many choices of $[\ell]$ and gauge choices for the $U(1)$ gauge
so that these requirements are satisfied. Further, given such choices, 
$\kappa_\ell$ and $\Phi_\ell \equiv - \ell^\az \cA_\az$ are constants 
over $\Delta$.
However, even with the equivalence class chosen it is easy to see from the 
transformation law given above that the value of $\kappa_\ell$ isn't fixed. 
A specific $\ell^\az \in [ \ell ]$ must be chosen to fix that constant over 
$\Delta$. Similarly, 
enough gauge freedom remains so that $\Phi_\ell$ may be assigned any value
that is desired (though once chosen that value is fixed 
everywhere on $\Delta$).

\begin{defn}
A {\bf rigidly rotating horizon (RRH)} $(\Delta,[\ell],\varphi)$, is a 
weakly isolated horizon with rotational symmetry. That is, 
$\mathcal{L}_\varphi \ell^\az = 0$, 
$\mathcal{L}_\varphi q_{\az \bz} = 0$,
$\mathcal{L}_\varphi \omega_\az = 0$, 
$\mathcal{L}_\varphi (q_\az^\bz \cA_\bz)=0$,
and $\mathcal{L}_\varphi (q_\az^\cz q_\bz^\dz F_{\cz \dz})=0$. 
Further $\varphi^\az$ has closed, 
circular orbits, and is normalized so that those 
orbits have affine length $2 \pi$. 
\end{defn} 

We will say that a specific foliation $\Delta_t$ is {\em adapted} to 
an RRH if $\varphi^\az \in T\Delta_t$ for all $t$ surfaces. 
Correspondingly, the time flow field is {\em adapted} if 
$T^\az = \ell'^\az - \Omega_\varphi \varphi^\az$ for some 
$\ell'^\az \in [\ell^\az]$ and some real constant $\Omega_\varphi$. 
Then, an adapted flow of time 
selects a specific ``normalization'' of $\ell^\az$ and therefore
an acceleration $\kappa_\ell$ as well. 
Equivalently, we can say that a choice of $\kappa_\ell$ and
$\Omega_\varphi$ selects a $T^\az$ on $\Delta$. In that case
$T^\az$ is no longer an independent vector field but is instead
defined with respect to quantities intrinsic to the 
surface\footnote{This may set off warning signals for readers who are
familiar with metric formulations of the action principle. In those cases,
$\Sigma_t$ and $T^\az$ are taken as background structures independent of the
spacetime variables and therefore $\delta T^\az = 0$. Thus one might
worry that this change will disrupt the variational 
calculations of the next section. This is not the case as they are based
on Hamiltonian rather than Lagrangian variations, and so don't assume
that $\delta T^\az = 0$. As far as the Hamiltonian is concerned
$T^\az$ doesn't exist, so whether it is varied or not is a matter of 
indifference}.

The canonical example of a rigidly rotating horizon is, of course, the
event horizon found in Kerr-Newman spacetime. Any linear combination of
the timelike and angular Killing vector fields of the spacetime is then
an adapted flow of time. Keep in mind  however, that the $t=\mbox{constant}$ 
surfaces in the standard Boyer-Lindquist coordinates are not spacelike 
everywhere and therefore not a suitable foliation from the current perspective.
Instead one should think in terms of the Painlev\'{e}-Gullstand-type 
coordinates defined in \cite{PGkerr}.

Finally, for completeness we note that an {\bf isolated horizon} is a weakly 
isolated horizon on which $\mathcal{L}_\ell$ and $\triangle_\az$ commute. We
won't actually make use of these in this paper, though often will follow
convention and use ``isolated horizon'' as a generic term for any of the 
surfaces discussed above. 

\subsubsection{Constant quantities on the horizons}
\label{const}

Given a non-expanding horizon $\Delta$ which is foliated by 
$\{\Delta_t : t_1 \leq t \leq t_2\}$ one can define an area as well as the 
electric and magnetic charges of each surface $\Delta_t$:
\bea
A_{\Delta}(t) &\equiv& \int_{\Delta_t} d^2 x \ssg, \\
Q_{\Delta}(t) &\equiv& \int_{\Delta_t} d^2 x \ssg E_\vdash, \mbox{ and}\\
P_{\Delta}(t) &\equiv& \int_{\Delta_t} d^2 x \ssg B_\vdash.
\eea
In the above, $E_\vdash \equiv (1/4 \pi) F_{\az \bz} \ell^\az \tl^\bz = 
(1/4 \pi) F_{\az \bz} u^\az n^\az$ is the normal component of the electric 
field to $\Delta_t$, while  $B_\vdash \equiv - (1/4 \pi) 
F^\star_{\az \bz} \ell^\az \tl^\bz = - (1/4 \pi) F^\star_{\az \bz} 
u^\az n^\az$ is the normal component of the magnetic field. 
$F^\star_{\az \bz}$ is the dual electromagnetic field tensor. Note too that 
though we have chosen to write $Q_{\Delta_t}$ and $P_{\Delta_t}$ in terms
of the normals and field tensors, they could also have been written purely as 
integrals of the pull-backs of the field tensors into $\Delta_t$ 
in which case their independence of the normals is manifest.

Now, it is quite easy to show that if the $\Delta_t$ are all
closed surfaces (which they are by construction), then these quantities
are constant in coordinate time and therefore not dependent on which 
surface $\Delta_t$ they are evaluated on. Therefore, the area and electric 
and magnetic charges of a non-expanding horizon are invariants of 
$\Delta$ and independent of the choice of foliation (though of course we
have already set $P_\Delta = 0$ with our assumption that a single $\cA_\az$ 
covers $\cM$). 

Next, moving up a step to weakly isolated horizons and continuing to assume 
that the $\Delta_t$ are closed, one can show that for any constants $c_1$ and 
$c_2$ and vector field $v^\az \in T \Delta$ such that $\pounds_\ell v^\az = 0$,
\bea
C_{\Delta}(v) 
&\equiv& \int_{\Delta_t} d^2 x \ssg v^\az (c_1 \omega_\az + c_2 E_\vdash 
\cA_\az),
\eea 
is a constant in coordinate time, and therefore an invariant of the horizon.
In particular, for a rigidly rotating horizon we can define the invariant 
\be
J_\Delta 
\equiv 
- \int_{\Delta_t} d^2 x \ssg \varphi^\az (\omega_\az/(8 \pi)  
- E_\vdash \cA_\az),
\ee
which in section \ref{phys} will be seen to be the angular momentum of
the horizon. 

Thus for a rigidly rotating horizon we have four geometrically defined
invariants: $A_\Delta$, $J_\Delta$, $Q_\Delta$, $P_\Delta$ (though we have
a priori assumed that the magnetic charge is zero).
$\kappa_\ell$ and $\Omega_\varphi$ are also constants over the surface
if we single out a specific adapted flow of time.
Later we will see that in order for the Hamiltonian formalism
to be well defined, these constants (along with $\Phi_\ell$) 
will have to be functions of $A_\Delta$, $J_\Delta$, and $Q_\Delta$
(though there will be quite a bit of freedom in defining the functions).
Therefore, $T^\az$ will be a ``live'' vector field and dependent on the 
spacetime over which it is defined \cite{ABL}.

Finally, for a given adapted time flow, 
\bea
\label{EDelta}
E_\Delta(T) &\equiv&
\int_{\Delta_t} d^2 x \ssg T^\az \left(\omega_\az/(4 \pi)  
- E_\vdash (2 \cA_\az - \Phi_\ell \tl_\az) \right) \\
&=& \frac{\kappa_{\ell}}{4 \pi} A_{\Delta}
+ 2 \Omega_\varphi J_{\Delta} + \Phi_{\ell} Q_{\Delta}  \nn
\eea
is also an invariant. If $E_\Delta(T)$ could be interpreted as the energy
of an RRH then this would be a Smarr relation for RRHs connecting their
energy, angular momentum, and electric charge. In fact, in section  
\ref{phys} we will see that $E_\Delta(T)$ is very closely related to the 
Komar/Bondi energy for a stationary spacetime and so can be taken as the  
the total energy of the RRH.

With this review under our belts and promise for the future in our minds, 
we turn examine the quasilocal Hamiltonian formulation for general relativity
in the presence of null boundaries.

\section{The metric-based Hamiltonian}
\label{calcs}
Over a four-manifold $\cM$ with no boundary, the standard action for 
gravity coupled to electromagnetism is
\be
\label{EinHil}
I_{EM} = \frac{1}{16 \pi } \int_{\cM} d^4 x \sqrt{-g} (\cR-\Lambda 
- F_{\az \bz} F^{\az \bz}).
\ee 
$\cR$ is the four-dimensional Ricci curvature
scalar, $\Lambda$ is the cosmological constant, and as defined earlier,
$g_{\az \bz}$ and $F_{\az \bz}$ are the metric and Maxwell tensors. 
Taking the variation of this action with respect to the metric and 
solving $\delta I = 0$ one recovers the standard equations of motion.

If, however, one considers a manifold $M$ with boundaries, 
then boundary conditions must be enforced on $\partial M$ in
order to keep the first variation well-defined. It is well known that these
conditions encode themselves in the variational principle 
in the form of boundary terms added to the standard action (see for example
\cite{BY1} for a discussion of this point). These boundary terms are well
understood for timelike and spacelike boundaries but not so well studied
if a boundary is null. Thus, an important part of the work of this 
paper will be to derive the appropriate boundary conditions for such 
null boundaries. This is done by calculating the variation of 
(\ref{EinHil}) including the effects of the boundaries and then reading off
which quantities must be fixed to keep the variation well defined. 
If necessary boundary terms may be added to change which quantities are to be
fixed. 

However, at the same time, we are interested in working within a 
Hamiltonian rather than Lagrangian framework. Thus, we want to take the 
Legendre transform of $I_{EM}$ to get the corresponding Hamiltonian. The first
step to taking that transformation is to rewrite the action in terms of 
quantities defined on the three-dimensional hypersurfaces rather 
than purely geometrical quantities on a four-manifold. To that end, and 
assuming the full set of boundaries defined in section \ref{SetUp4D} 
(and where for now $\Delta$ is just 
a null surface rather than an isolated horizon of one type or another), 
we may rewrite equation (\ref{EinHil}) as
\bea
I_{EM} &=&  - \int_{\Sigma_{21}} d^3 x (P^{\az \bz} h_{\az \bz}) 
- \int_{\Delta_{21}} d^2 x (P^\Delta_\ssg \ssg)
- \int_{B_{21}} d^2 x (P^B_\ssg \ssg) 
\label{3daction} \\
&& + \int dt \left\{ 
\int_{\Sigma_t} d^3 x (P^{\az \bz} \mathcal{L}_T h_{\az \bz} + \cE^\az
\mathcal{L}_T \mathcal{A}_\az) \right\} \nn \\
&& + \int dt \left\{ \int_{\Delta_t} d^2 x 
(P^\Delta_\ssg \mathcal{L}_T \sqrt{\sigma})
+ \int_{B_t} d^2 x (P^B_\ssg \mathcal{L}_T \sqrt{\sigma}) \right\} 
\nn \\
&& + \int dt \left\{ 
- \int_{\Delta_t} d^2 x T^\az (\omega_\az/( 8 \pi)  
+ E_\vdash \mathcal{A}_\az) 
+ \int_{B_t} d^2 x \ssg T^\az (  \bu^\cz \nabla_\az 
\bn_\cz /(8 \pi) + E_\vdash \cA_\az) 
\right\} \nn \\
&&  - \int dt H_{blk}
\nn
\end{eqnarray} 
where $\int_{\Sigma_{21}} = \int_{\Sigma_2} - \int_{\Sigma_1}$, 
$\int_{\Delta_{21}} = \int_{\Delta_2} - \int_{\Delta_1}$, and
$\int_{B_{21}} = \int_{B_2} - \int_{B_1}$. 
$P^\Delta_\ssg \equiv 1/(8 \pi ) \log \xi$ and
$P^B_\ssg \equiv -1/(8 \pi ) \sinh^{-1} \eta$ will be seen
to be momenta conjugate to $\ssg$ on $\Delta_t$ and $B_t$, 
while $P^{\az \bz}$ and $\cE^\az$ are the 
hypersurface momentum and densitized electric field conjugate to 
$h_{\az \bz}$ and $A_\az$ respectively.  
Further,
\be
H_{blk} = \int_{\Sigma_t} d^3 x 
(N \cH + V^\az \cH_\az + T^\az A_\az \cQ)
\ee
where $\cH \equiv - \sqrt{h}/(8 \pi ) (G_{\az \bz} + \Lambda g_{\az \bz} - 
8 \pi  T_{\az \bz}) u^\az u^\bz$, and
$\cH_\az \equiv 
- \sqrt{h}/(8 \pi ) h_\az^\cz (G_{\cz \bz} + \Lambda g_{\cz \bz} - 
8 \pi  T_{\cz \bz}) u^\bz$
are the standard energy and momentum constraint equations for spacelike
hypersurfaces, while $\cQ \equiv - D_\az \cE^\az$ is the Gauss constraint
equation.

From a Hamiltonian perspective each of these quantities may be recast
as defined entirely on $\Sigma_t$. Then, with some prescience of how the
calculations will come out, and which boundary terms we will want to fix, 
we define the new action, 
\bea
I &\equiv&  I_{EM} + \int_{\Sigma_{12}} d^3 x (P^{\az \bz} h_{\az \bz})
+ \int_{\Delta_{21}} d^2 x (P^\Delta_\ssg \ssg) +  
\int_{B_{21}} d^2 x (P^B_\ssg \ssg) \label{blob1} \\
&& + \int_\Delta d^3 x  \ssg T^\az (\omega_\az/( 8 \pi)  
+ E_\vdash \mathcal{A}_\az) 
- \int_{B_t} d^3 x \sqrt{- \gamma} \Theta, \nn
\eea
where $\Theta\equiv - \gamma^{\az \bz} \nabla_\az \bn_\bz$ 
is the extrinsic curvature of $B$ in $M$. With the exception of the term at
$\Delta$, these are the standard terms 
added to a quasilocal action to ensure that its first variation is 
well-defined when the boundary metrics (and potentials) are held constant
\cite{hayward, naked}. They are independent of the time flow, 
and so the action with these boundary terms remains similarly independent.
By contrast, the term on $\Delta$ explicitly incorporates the time flow. 
As we shall soon see, this inclusion means that the Hamiltonian 
formalism places restrictions on the allowed forms of $T^\az$. This is in 
marked contrast to the traditional formulation where it is an independent
background structure. That said, the new action may be rewritten as a 
coordinate-time-integrated functional of quantities defined entirely in 
$\Sigma_t$:
\bea
I &=& \int dt \left\{ 
\int_{\Sigma_t} d^3 x (P^{ab} \dot{h}_{ab} + \cE^a \dot{A}_a) 
+ \int_{\Delta_t} d^2 x (P^\Delta_\ssg \dot{\sqrt{\sigma}})
+ \int_{B_t} d^2 x (P^B_\ssg \dot{\sqrt{\sigma}}) \right\} 
\label{action} \\
&& - \int dt \left\{ H_{blk} + \int_{B_t} d^2 x \ssg
(\bN (\bep+\bep^m) - \hat{V}^a (\bj_a + \bj_a) )
\right\}. \nn
\eea
We are now adopting a Hamiltonian point of view, and so the interpretation of
many of the quantities in the above expression has changed. First and foremost,
the momenta $P^{ab}$, $\cE^a$, $P^\Delta_\ssg$, and $P^B_\ssg$ are now 
independent fields with no {\em a priori} connection to their conjugate 
configuration variables. In particular, the time derivatives of those
configuration variables (indicated by dots) are no longer geometrically
defined by Lie derivatives in the $T^\az$ direction. From the
three-dimensional perspective this direction no longer exists and the
time derivatives are now freely determined as well 
(though of course the variational
principal will ultimately link them back to the conjugate momenta). 
The boundary terms on $B_t$ are well-known from the work of Brown-York
\cite{BY1} and its non-orthogonal generalizations \cite{nopaper, naked}. 
In three-dimensional language:
\bea
\bep &\equiv& k/(8 \pi \lambda) + 2 \eta P^{ab} n_a n_b /\sqrt{h}, \\
\bep^m &\equiv& -(\cE^b n_b/\sqrt{h})(\Phi/\lambda - \eta A_a n^a),\\
\bj_a &\equiv& -
2 \sigma_{ac} P^{cd} n_d /\sqrt{h} - \lambda/(8 \pi) d_a \eta, 
\mbox{ and} \\
\bj^m_a &\equiv&  -(\cE^b n_b/\sqrt{h}) \sigma_a^b A_b.
\eea
Their meaning in more clear from a four-dimensional perspective. There, 
$\bep = \bar{k}/(8 \pi)$ where $\bar{k}$ is the extrinsic curvature of 
$B_t$ with respect to the normal $\bn^\az$, $\bep^m = E_\vdash \bar{\Phi}$
where $\bar{\Phi} = - \bu^\az \cA_\az$ is the Coulomb potential with respect
to $\bu^\az$, $\bj_\az = 1/(8 \pi) \sigma_\az^\bz \bu^\cz \nabla_\bz \bn_\cz$,
and $\bj^m_\az = E_\vdash \sigma_\az^\bz \cA_\bz$. Physically, the $\bep$ 
terms are related to energy while the $\bj_a$ terms are connected with 
angular momentum \cite{BY1}. 

Applying the Legendre transform, we identify the terms on the first
line of (\ref{action}) 
as kinetic-energy-type terms and the last line as the negative of the 
Hamiltonian $H$ (in standard classical mechanics the analogous 
relationship is $I = p \dot{q} - H$).
The Hamiltonian equations of motion and the corresponding boundary 
conditions are found by solving $\delta I = 0$ where the action is
in its dynamical three-surface form given in equation (\ref{action}).
The only really 
troublesome part of the calculation is finding the variation of $H_{blk}$.
Luckily, however, its variation has previously been calculated in
\cite{mythesis} and for the pure gravitational case (using slightly 
different notation) in \cite{BYLnonortho}. Then, varying with
respect to $h_{ab}$, $A_a$, $P^{ab}$, $\cE^a$, $N$, and $V^a$ we have
\begin{eqnarray} 
\delta H_{blk} &=& 
\int_{\Sigma_t} d^3 x 
\left\{ (\mathcal{H} - \Phi \cQ) \delta N 
+ (\mathcal{H}_a - A_a \cQ) \delta V^a 
- N \cQ \delta \Phi) \right\} \\
&& + \int_{\Sigma_t} d^3 x \left\{  
 [h_{ab}]_T \delta P^{ab} - [P^{ab}]_T \delta h_{ab}
+ [A_a]_T \delta \cE^a - [\cE^a]_T \delta A_a 
\right\} \nn \\ 
&& + \int_{B_t - \Delta_t} d^2 x 
\left\{ - N \left( \delta [ \ssg \varepsilon ] 
+ v \delta [ \sqrt{\sigma} \varepsilon^\updownarrow ]
\right)  + \hat{V}^a \delta [ \sqrt{\sigma} \hat{\jmath}_a ] 
- \frac{N \sqrt{\sigma}}{2}
\left( s^{ab} + v s^{\updownarrow ab} \right) \delta
 \sigma_{ab} \right\} \nonumber \\
&& + \int_{B_t - \Delta_t} d^2 x \left\{ 
(N(\Phi + v n^a A_a) 
\delta [ (\ssg/\sqrt{h}) \cE^a n_a ]  
- V^a \delta [ (\ssg/\sqrt{h}) \cE^b n_b \hat{A}_a ] 
\right\} \nn \\
&& + \int_{B_t - \Delta_t} d^2 x \left\{ 
N (\ssg/\sqrt{h}) (\hat{\cE}^a + v n_c \epsilon^{cab} \hat{\cB}_b) \delta \hat{A}_a
\right\},
\nn
\end{eqnarray} 
where 
$[h_{ab}]_T$, $[P^{ab}]_T$, $[A_a]_T$, and $[\cE^a]_T$ are each equal to
the corresponding Lie derivatives in the direction $T^\az$ (though of course
they have, in this case, been obtained from the variational rather than 
a geometrical calculation). 
$\varepsilon = k/(8 \pi)$ (the energy
density on $B_t$ \cite{BY1}),
$\varepsilon^\updownarrow =  2/\sqrt{h} P^{ab} n_a n_b$ (the time rate 
of change of $\ssg$ \cite{naked}), and 
$\hat{j}_a = -2 \sigma_{ab} P^{bc} n_c/\sqrt{h}$
(the angular momentum density on $B_t$ \cite{BY1}). 
$v = V^a n_a /N$ (and can be thought of as 
the ``radial speed'' of the boundary relative to $\Sigma_t$), 
$\hat{V}^a = \sigma^a_b V^b$,
$s^{ab} = (1/8 \pi ) \left( k^{ab} - (k - n^c a_c) \sigma^{ab} \right)$
(the stress tensor on $B_t$ induced by the gravitational field 
\cite{BY1}), 
$a_a = (1/N) D_a N$ (which from a four-dimensional perspective is 
$a_\az = u^\bz \nabla_\bz u_\az$ -- the acceleration of the
timelike normal), $s^{\updownarrow ab} = (1/8 \pi )
\left(l^{ab} - (l + u^\cz a^{\updownarrow}_\cz) \sigma^{ab} \right)$ 
(the ``time'' stress tensor \cite{BYLnonortho}), and 
$a^{\updownarrow}_\az = n^\bz \nabla_\bz n_\az$ (the acceleration of the
spacelike normal along its length). $\epsilon^{abc}$ is the
Levi-Cevita tensor on $\Sigma_t$ and $\cB^a = \sqrt{h}/(4 \pi ) 
\epsilon^{abc} D_b A_c$ is the densitized magnetic field.

Now, if $-1<v<1$ then the evolution of the boundary $\Omega$ is timelike, 
the case that was considered in great detail in \cite{nopaper, naked, 
mythesis, BYLnonortho} (where the reader is directed for further details on
the result that we will quote below for $B_t$).
On the other hand if $v= \pm 1$ then the evolution
is null and we are considering a situation which has not been directly 
investigated until now (though it was considered as a limiting case in 
\cite{BYBondi} and \cite{cadoni}). 
We will set $v=1$ ($v=-1$ can then be considered simply by 
reversing the sign of $n_a$). Thus the boundary $\Omega$ expands/contracts 
with the speed of light in the direction $n^a$ (though for the non-expanding
horizons that we consider that actually means it isn't expanding at all). 

Then harking back to the definitions surrounding equation 
(\ref{thetaab}), one can write
\bea
\theta_{ab} &=& -\xi \left(k_{ab} + l_{ab}\right) \label{31}\\
\hat{\omega}_a &=& - \sigma_a^b n^c K_{bc} + 8 \pi  d_a P^\Delta_\ssg 
\label{32}\\
\kappa_\ell &=&  \xi \left( \frac{1}{N} n^a D_a N - n^a n^b K_{ab} 
+ \frac{8 \pi }{N} (\dot{P}_\ssg^\Delta - \mathcal{L}_{\hat{V}} 
P_\ssg^\Delta)
\right) \label{33}
\eea
where one recognizes that in the Hamiltonian context,
$\xi = \exp( 8 \pi  P_\ssg^\Delta$), $K_{ab} = \frac{8 \pi }{\sqrt{h}} 
(P h_{ab} - 2 P_{ab} )$, and $l_{ab} = \sigma_a^c \sigma_b^d K_{cd}$. 
With all of this is mind (and with a quick consulation with 
\cite{mythesis} for the details on $B$) we can write $\delta H$ (recall
that this is the negative of the last line of equation (\ref{3daction})) as:
\begin{eqnarray} 
\delta H &=& 
\int_{\Sigma_t} d^3 x 
\left\{ (\mathcal{H} - \Phi \cQ) \delta N
+ (\mathcal{H}_a - A_a \cQ) \delta V^a 
- N \cQ \delta \Phi) \right\} \\
&& + \int_{\Sigma_t} d^3 x \left\{  
 [h_{ab}]_T \delta P^{ab} - [P^{ab}]_T \delta h_{ab}
+ [A_a]_T \delta \cE^a - [\cE^a]_T \delta A_a 
\right\} \nn \\ 
&& + \int_{\Delta_t} d^2 x 
\left\{ - \tN \delta[ \ssg \theta/(8\pi )] 
        - \hat{V}^a \delta [ \ssg \hat{\omega}_a/(8 \pi )]
        - (\tN/2) \tilde{s}^{ab} \delta \sigma_{ab} \right\} \nn \\
&& + \int_{\Delta_t} d^2 x \left\{ 
\tN \Phi_\ell \delta[(\ssg/\sqrt{h}) n_b \cE^b] - 
\ssg/(4 \pi ) \tN f^a \delta A_a - \hat{V}^a 
\delta [ (\ssg/\sqrt{h}) n_b \cE^b \hat{A}_a ] \right\} \nn \\
&& + \int_{B_t} d^2 x \ssg 
\left\{ (\bep+\bep^m) \delta \bN - (\bj_a + \bj^m_a) \delta V^a 
- (\bN/2) \bs^{ab} \delta \sigma_{ab} \right\} \nonumber \\
&& + \int_{B_t} d^2 x \tN \ssg /\sqrt{h}
\left\{ [\cE^a n_a] \delta \bPhi + \bar{\cB}_a n_b \epsilon^{abc} 
\delta \hat{A}_c \right\} \nn \\
&& + \int_{\Delta_t} d^2 x \left\{ [\ssg]^\Delta_T \delta
P^\Delta_\ssg - [ P^\Delta_\ssg ]_T \delta \ssg
\right\} \nonumber \\
&& + \int_{B_t} d^2 x \left\{ [\ssg]^B_T \delta
P_{\sqrt{\sigma}}^B - [ P^B_\ssg ]_T \delta \sqrt{\sigma}
\right\} \nonumber, 
\end{eqnarray} 
where of the newly appearing quantities,
$\tilde{s}^{ab} = \frac{1}{8 \pi } (\theta^{ab} + [\kappa_\ell - \theta] 
\sigma^{ab})$,  
$\bs^{ab} = \frac{1}{8 \pi } (\bk^{ab} + [\ba_c \bn^c - \bk] 
\sigma^{ab})$, $\bar{\cB}^b = \sqrt{h}/(4 \pi ) 
\epsilon^{abcd} \bu_a \nabla_b A_c$,
$\left[ \sqrt{\sigma} \right]^\Delta_T = - \tilde{N}
\ssg \theta + \mathcal{L}_{\hat{V}} \ssg$,
$\left[ \sqrt{\sigma} \right]^B_T =
- \bN \ssg \bar{l} + \mathcal{L}_{\hat{V}} \ssg$, 
and $[ P_{\sqrt{\sigma}}^0]^\Delta_T$ and $[ P_{\sqrt{\sigma}}^0]^B_T$ are
freely defined functions.

The Hamiltonian equations of motion and attendant boundary 
conditions are then be found by solving $\delta I = 0$. We have:
\bea
\delta I &=& 
\mbox{(Terms that vanish for solutions to the equations of motion)} 
\label{Act2Var}\\
&& + \int_\Sigma d^3 x (P^{ab} \delta h_{ab} + \cE^a \delta A_a) + 
\int_{\Delta_t} d^2 x P_\ssg^\Delta \delta \ssg
+ \int_{B_t} d^2 x P_\ssg^B \delta \ssg \nn \\
&& - \int dt \int_{\Delta_t} d^2 x 
\left\{ \tN \delta[ \ssg \theta/(8 \pi )] 
        +\hat{V}^a \delta [ \ssg \hat{\omega}_a /(8 \pi )]
        +\frac{\tN \ssg}{2} \tilde{s}^{ab} \delta \sigma_{ab} \right\} \nn \\
&& - \int dt \int_{\Delta_t} d^2 x \left\{ 
\tN \Phi_\ell \delta[(\ssg/\sqrt{h}) n_b \cE^b] - 
\ssg/(4 \pi ) \tN f^a \delta A_a - \hat{V}^a 
\delta [ (\ssg/\sqrt{h}) n_b \cE^b \hat{A}_a ] \right\} \nn \\
&& - \int dt \int_{B_t} d^2 x \ssg 
\left\{ (\bep+\bep^m) \delta \bN - (\bj_a + \bj^m_a) \delta V^a 
- (\bN/2) \bs^{ab} \delta \sigma_{ab} \right\} \nonumber \\
&& - \int dt \int_{B_t} d^2 x \tN \ssg /\sqrt{h}
\left\{ [\cE^a n_a] \delta \bPhi + \bar{\cB}_a \bn_b \epsilon^{abc} \delta 
\hat{A}_c \right\}, \nn
\eea
where to tidy up an already formidable expression, we haven't written
out the equations of motion term. It only goes to zero though if
the Einstein-Maxwell constraints $\cH = 0$, $\cH_a=0$, and $\cQ = 0$ are
satisfied, along with the time-evolution equations
$\dot{P}^{ab} = [P^{ab}]_T$ and $\dot{\cE}^a = [\cE^a]_T$ (which encode
the Einstein-Maxwell evolution equations), and 
$\dot{h}_{ab} = [h_{ab}]_T$, $\dot{A}_a = [A_a]_T$ 
and $\dot{\ssg} = [\ssg]_T$ (which 
establish the connection between the conjugate momenta and
the time derivatives of the configuration 
variables). Thus, the full four-dimensional structure and 
equations of motion come from the three-dimensional action and solving
$\delta I = 0$. 

These equations of motion are, of course, well understood and nothing new.
It is more interesting to examine the boundary conditions that must
be imposed on solutions so that the variation will be well defined. In the
first place, by the second line of (\ref{Act2Var}), $h_{ab}$ and $A_a$ should 
be fixed on the initial and final surfaces. This fixing of initial and final
conditions is, of course, standard for a variational principle. 

In the meantime, at each surface $B_t$, the metric, lapse, shift, 
Coulomb potential, and tangential components of the vector potential should 
all be prespecified as boundary conditions. Note however, that fixing these
quantities does not fix any thermodynamic quantities of interest. As shown 
by Brown and York, the energy and angular momentum both depend on 
$\nabla_\az \bn_\az$ which is not determined by $\gamma_{\az \bz}$. 
At the same time, the electromagnetic conditions are sufficient only 
to fix the magnetic charge, which we have assumed to be zero anyway. The
electric charge is left free. Thus, with these boundary condititions, 
the variation is well defined but the allowed thermodynamic quantities on 
$B$ are not fixed. Note too that for any function/functional $I_0$ of the 
induced metric $\gamma_{\az \bz}$ and vector potential 
$\gamma_\az^\bz \cA_\bz$, $\delta I_0 = 0$. Thus, any such term may be added 
onto the action without affecting the equations of motion or even variational
boundary terms. This is the source of the zero-point energy ambiguity discussed
in the introduction. The obvious choice is to set $I_0$ to zero and then
forget about it, but unfortunately for such a choice the action is non-zero
and even diverges for infinite regions of Minkowski space, let alone any 
curved spacetime. Thus, $I_0$ should be chosen to compensate for this 
problem. A discussion of the various choices that are made may be found
in \cite{naked, mythesis}, but here we just need to keep in mind that 
the choice is not determined by the formalism. 

Now, let us turn our attention to $\Delta$. If it is an RRH with
an (unspecified) adapted flow of time, we can use the properties derived in 
section \ref{IHreview} to show that the boundary term at $\Delta$ becomes:
\be
\delta I_\Delta = - \int dt \left(
\frac{\kappa_\ell}{8 \pi } \delta A_\Delta
           +\Phi_\ell \delta Q_{\Delta} 
            +\Omega_\varphi \delta J_\Delta \right) \label{f1}.
\ee
Since all of the quantities are constant in time, the integral becomes a
multiplicative factor of $(t_2-t_1)$.

Now, a simple but at the same time slightly subtle point must be made. 
For a given isolated horizon, we have seen 
that $A_\Delta$, $J_\Delta$, and $Q_\Delta$ are constants and so at
first blush one might think that the above term is automatically zero, 
as for the corresponding the situation on $B$ where the fixed boundary 
metric sets the
variational boundary term to zero. On $\Delta$ 
however, we haven't specified that it
is a {\em specific} RRH, 
but only that it is  rigidly rotating horizon.
Therefore, the variation considered includes the freedom to move between
different isolated horizon solutions and consequently different
values of those constants. Consequently, these terms are not constant with
respect to the variation. Thus, as things stand, $\delta I$ 
is not zero for variations between these
perfectly well-defined solutions to the equations of motion. However, 
this can be fixed as there is still the freedom to 
add a boundary term $I_\Delta$ on $\Delta$ (just as a boundary
term has already been added at $B$ in equation (\ref{action}) in order
to fix the boundary metric and ensure that no thermodynamic quantities
have been inadvertently fixed). Then, $I_\Delta$ must take the form
\be
I_\Delta = - \int dt H_\Delta(A_\Delta, Q_\Delta, J_\Delta), 
\ee
where $H_\Delta$ must satisfy the relations
\be
\label{pfl}
\frac{ \partial H_\Delta}{\partial A_\Delta} = \frac{\kappa_\ell}{8 \pi},
\hspace{.5cm}
\frac{ \partial H_\Delta}{\partial J_\Delta} = \Omega_\varphi,
\hspace{.25cm} \mbox{and} \hspace{.25cm} 
\frac{ \partial H_\Delta}{\partial Q_\Delta} = \Phi_\ell.
\ee
This set of equations only makes sense if $\kappa_\ell$, 
$\Phi_\ell$, and $\Omega_\varphi$ are themselves functions of
$A_\Delta$, $Q_\Delta$, and $J_\Delta$ though the form of their 
dependence is not fully determined by $\delta I = 0$.
As has already been seen, the choice of a $\kappa_\ell$ and
a $\Omega_\varphi$ is equivalent to a choice of $T^\az$ on the boundary
$\Delta$. Thus, the allowed $T^\az$ are now restricted by
the spacetime configuration in contrast to the usual quasilocal formulation 
where the $T^\az$ is an independent background structure that may be freely 
chosen. For a discussion of this from the 
symplectic point of view, the reader is referred to \cite{AFK, ABL}.

\section{Physical quantities and thermodynamics}
\label{phys}

We now turn to a physical interpretation of this mathematics.

\subsection{The first law of RRH mechanics}

To make the physical connection with thermodynamics, let us restrict
ourselves to the space of all (magnetic-charge free) 
solutions to the Einstein-Maxwell equations that have an
inner boundary $\Delta$ that is a RRH and an outer boundary $B$ which 
has a given induced metric $\gamma_{\az \bz}$ and vector potential 
$\gamma_\az^\bz \cA_\bz$. Further, we will assume that $B$ has a timelike
Killing vector field and so its physical fields are stationary. As 
discussed in \cite{BY1} this means that the value of the Hamiltonian
boundary term evaluated on $B$ is time (and therefore exact 
$B_t$ surface) independent. Then, the value of the Hamiltonian 
for any element of this space is given by
\be
H_{sol} \equiv - H_\Delta(A_\Delta,Q_\Delta,J_\Delta) 
+ \int_{B_t} d^2 x \ssg
\left(\bN (\bep+\bep^m - \bep^0) - \hat{V}^a (\bj_a + \bj^m_a) \right),
\ee
though it doesn't matter which surface $B_t$ is used for the
evaluation of the outer boundary term. We may write the outer boundary term
as the functional $H_B[\bn^\az,\cA_\az \bn^\az]$ since the only freedom 
left to it comes from $\bn^\az$ and $\bn^\az \cA_\az$. $\bep^0$ is an 
extra boundary term arising from the reference term $I_0$. Its exact form
doesn't matter here, though keep in mind that it is usually chosen so that 
$H_{sol}$ (without
the inner boundary terms) matches the ADM energy in the appropriate limit
for asymptotically flat spacetimes. 

In spite of the above restrictions, the phase space under consideration
is still extremely large. To wit, the inner boundary term demands that
$\Delta$ be an RRH, but does not require that it be any particular RRH. 
The outer boundary term fixes the induced metric and vector potential
on $B$, but does not place restrictions on the other components of those
quantities. Thus even though the outer boundary
term defines the ADM mass if $\cM$ is asymptotically flat and 
$B$ is taken to timelike infinity \cite{BY1}, 
the variations are such that the value of that mass is not fixed. 
In the same way, the total electric charge and angular momentum of the 
spacetime are not fixed in that limit. Similarly, as was discussed in the
last section, the 
quasilocal quantities are undetermined even if $B$ is finite.
Further, though the assumptions mean that matter/energy is not allowed to 
flow across either boundary, there is still a large amount of freedom
left to such flows inside $M$. A study of the range of spacetimes containing 
isolated horizons may be found in \cite{geometry}.

Now given that $H_B$ corresponds to the CQLE/ADM mass inside $B$ it is natural
to interpret $H_\Delta$ as the energy of the RRH. As it stands though,
this is not very satisfactory since there
is so much freedom left in its
definition. This problem will be addressed in the next section where we'll 
investigate the Kerr-Newman region of the phase space of solutions. For 
now though, let's tentatively make this association and see where it leads.

With respect to variations through the phase space of solutions, we know
from equations (\ref{pfl}) that
\be
\label{firstlaw}
\delta H_\Delta = \frac{\kappa_\ell}{8 \pi } \delta A_\Delta + 
\Omega_\varphi \delta J_\Delta + \Phi_\ell \delta Q_\Delta. \nn
\ee
Then, interpreting $\delta H$ as the energy of the RRH, $\kappa_\ell$
as the surface gravity/temperature, $J_\Delta$ as the angular momentum,
and $\Omega_{\varphi}$ as the angular velocity conjugate to $J_\Delta$, this
is the first law of thermodynamics. However, note that there
is still freedom in the definition of $\kappa_\ell$, 
$\Omega_{\varphi}$, and the Coulomb potential $\Phi_\ell$ and so at the 
moment this is just the first law of isolated horizon mechanics and it
is important to note that it holds for all of the possible forms of 
$\kappa_\ell$, $\Omega_{\varphi}$, and $\Phi_\ell$.
The next section will make the physical connection but for now we 
point out
that this law must hold for the Hamiltonian evolution to be consistent -
or conversely the Hamiltonian evolution gives rise to the first law. This
result was found via symplectic methods in \cite{AFK} and \cite{ABL}.

It is not hard to show that variation of the outer term in this case is
\be
\delta H_B = - H_B \bn^\az \delta \bn_\az,
\ee 
and we note that it is not coupled in any way to the variation on the inner 
boundary. This is not surprising given the freedom of the phase space. Energy
can be injected into the region $M$ without affecting the RRH itself, so there
is no reason why the inner and outer boundary energies should change in 
lock-step. An example of such a process would be to insert a 
spherically symmetric
shell of matter around a Schwarzschild black hole. Then, the metric inside
the shell would be not be affected while the ADM energy would certainly 
increase. More complicated and dynamic situations could also be considered. 

\subsection{Physical interpretation and calibration}

The quickest way to extract the physical content of the law and the quantities
that it connects is to consider the Komar expressions for angular momentum and
energy. First, given a stationary spacetime that contains an RRH and has
a global rotational Killing vector field $\varphi^\az$, the Komar angular
momentum integral evaluated on a foliation element $\Delta_t$ is
\bea
J^K_\Delta &\equiv& - \frac{1}{16 \pi }
\int_{\Delta_t} d^2 x \ssg (\ell^\az \tl^\bz - \tl^\az \ell^\bz)
\nabla_\az \varphi_\bz \\
&=& \frac{1}{8 \pi }\int_{\Delta_t} d^2 x \ssg \varphi^\az \omega_\az, \nn
\eea
which of course is equal to the purely gravitational part of $J_\Delta$. 
At the same time it was seen in \cite{ABL} that the purely electromagnetic
part of $J_\Delta$ is equal to the total angular momentum
of the attendant electromagnetic fields. Specifically, if 
$B$ is stationed at infinity, $\cM$ is 
asymptotically flat, and $\varphi^\az$ is a global angular Killing 
vector, then for any surface $\Sigma_t$, one can show that
\be
J^K_\infty - J^K_\Delta 
= - \int_{\Sigma_t} d^3 x \sqrt{h} T_{\az \bz} u^\az \varphi^\bz 
= J^{EM}_\Delta
\ee
where $J_K^\infty$ is the Bondi mass, and $J^{EM}_\Delta$ is the 
electromagnetic component of $J_\Delta$. Therefore $J_\Delta = J_\infty$ 
for stationary spacetimes 
and so measuring $J_\Delta$
at the horizon, one gets not only the angular momentum of the hole itself but
also the angular momentum of all of its associated electromagnetic 
fields. This is, perhaps, a bit of a surprise since we have tacitly 
been working with the assumption that $J_\Delta$ measures the angular
momentum inside $\Delta_t$. On the other hand, given that the mathematics
has viewed $\Delta$ not as a boundary containing a black hole, but rather
as a boundary of $M$ perhaps this isn't such a surprise afterall.

Encouraged by this result, let us now consider the Komar integral for the 
energy. In that case, if an adapted time flow vector field $T^\az = \ell^\az 
- \Omega_\varphi \varphi^\az$ can be extended to a global Killing vector 
field, the
Komar integral for the energy can be evaluated on the horizon as
\bea
E^K_\Delta &\equiv& \frac{1}{8 \pi } \int_{\Delta_t} d^2 x \ssg 
(\ell^\az \tl^\bz - \ell^\bz \tl^\az) \nabla_\az T_\bz \\
&=& \frac{1}{4 \pi } \int_{\Delta_t} d^2 x \ssg T^\az \omega_\az \nn \\
&=&  \frac{\kappa}{4 \pi } A_\Delta + 2 \Omega_\varphi J^K_\Delta, \nn
\eea
which is equal to the purely gravitational part of the constant
$E_\Delta$ defined back in equation (\ref{EDelta}). Given the previous
result for angular momentum, it is tempting to guess that the remaining
electromagnetic part of $E_\Delta$ is equal to the stress-energy of the
surrounding electromagnetic fields. In fact this is the case, for if 
$M$ is an asymptotically flat stationary axisymmetric spacetime with global 
Killing vector field $T^\az$, then one can show that
\be
E^K_\infty - E^K_\Delta = 2 \int_{\Sigma_t} d^3 x T_{\az \bz} u^\az T^\bz
= E^{EM}_\Delta.
\ee
That said however, we have not yet shown
that $E_\Delta$ is a satisfactory $H_\Delta$. 
Let us now postulate just that, and see where it leads.

First, if we assume that $H_\Delta = E_\Delta$ then 
a Smarr-type formula holds for isolated horizons. Thus, let us instead
make the slightly more general assumption that
\be
H_\Delta = \frac{\kappa_\ell}{4 \pi} A_\Delta + 2 \Omega_\varphi J_\Delta 
+ \Phi_\ell Q_\Delta.
\ee
Any consequences of this assumption will then automatically hold for the 
special case where $H_\Delta = E_\Delta$ as well. 
Then, equations (\ref{pfl}) expand to a   
set of three coupled differential equations. Namely
\bea
0 &=& \frac{\kappa_\ell}{8 \pi } 
+ \frac{\partial \kappa_\ell}{\partial A_\Delta} \frac{A_\Delta}{4 \pi }
+ 2 \frac{\partial \Omega_\varphi}{\partial A_\Delta} J_\Delta
+ \frac{\partial \Phi_\ell}{\partial A_\Delta} Q_\Delta \label{e1}\\
0 &=& \Omega_\varphi
+ \frac{\partial \kappa_\ell}{\partial J_\Delta} \frac{A_\Delta}{4 \pi }
+ 2 \frac{\partial \Omega_\varphi}{\partial J_\Delta} J_\Delta
+ \frac{\partial \Phi_\ell}{\partial J_\Delta} Q_\Delta  \label{e2}\\
0 &=& \frac{\partial \kappa_\ell}{\partial Q_\Delta} \frac{A_\Delta}{4 \pi }
+ 2 \frac{\partial \Omega_\varphi}{\partial Q_\Delta} J_\Delta
+ \frac{\partial \Phi_\ell}{\partial Q_\Delta} Q_\Delta. \label{e3}
\eea
Quite a mess, but luckily these equations decouple 
when we consider the relationships between the partial derivatives of
$\kappa_\ell$, $\Omega_\varphi$, and $\Phi_\ell$ that may be derived from 
equations (\ref{pfl}). Partial derivatives of $H_\Delta$ in phase space
commute so 
\bea
\frac{1}{8 \pi } \frac{\partial \kappa_{\ell}}{\partial J_\Delta}
= \frac{\partial \Omega_\varphi}{\partial A_\Delta}, \hspace{.5cm}
\frac{1}{8 \pi } \frac{\partial \kappa_{\ell}}{\partial Q_\Delta}
= \frac{\partial \Phi_\ell}{\partial A_\Delta}, \hspace{.5cm} 
\frac{\partial \Omega_\varphi}{\partial Q_\Delta}
= \frac{\partial \Phi_\ell}{\partial J_\Delta},
\eea
and equations (\ref{e1}),(\ref{e2}), and (\ref{e3}) become
\bea
0 &=& \kappa_\ell 
+ 2 \frac{\partial \kappa_\ell}{\partial A_\Delta} A_\Delta 
+ 2 \frac{\partial \kappa_\ell}{\partial J_\Delta} J_\Delta
+ \frac{\partial \kappa_\ell}{\partial Q_\Delta} Q_\Delta,  \\
0 &=&  \Omega_\varphi 
+ 2 \frac{\partial \Omega_\varphi}{\partial A_\Delta} A_\Delta 
+ 2 \frac{\partial \Omega_\varphi}{\partial J_\Delta} J_\Delta
+ \frac{\partial \Omega_\varphi}{\partial Q_\Delta} Q_\Delta, \mbox{ and} \\
0 &=& 2 \frac{\partial \Phi_\ell}{\partial A_\Delta} A_\Delta 
+ 2 \frac{\partial \Phi_\ell}{\partial J_\Delta} J_\Delta
+ \frac{\partial \Phi_\ell}{\partial Q_\Delta} Q_\Delta.  
\eea
These have the general solutions:
\be
\kappa_\ell 
= \frac{1}{\sqrt{A}} f \left(
\frac{J_\Delta}{A_\Delta},\frac{Q_\Delta}{\sqrt{A_\Delta}} \right),
\hspace{.5cm}
\Omega_\varphi
= \frac{1}{\sqrt{A}} 
g \left(\frac{J_\Delta}{A_\Delta},\frac{Q_\Delta}{\sqrt{A_\Delta}}\right),
\hspace{.25cm} \mbox{and} \hspace{.25cm}
\Phi_\ell =  
h \left( \frac{J_\Delta}{A_\Delta},\frac{Q_\Delta}{\sqrt{A_\Delta}} \right),
\label{kdef} \ee
where $f(x,y)$, $g(x,y)$, and $h(x,y)$ are freely defined  functions on 
the real plane. Thus, given a Smarr-type formula for $H_\Delta$ (which 
in particular includes the case where $H_\Delta = E_\Delta$), then 
$\kappa_\ell$, $\Omega_\varphi$, and
$\Phi_\ell$ must have the functional dependences given above for the 
Hamiltonian evolution to be well-defined. In any such case the first
law of isolated horizon mechanics will hold. 

Now, that said, for at least one region of the RRH phase space there is a
natural choice for $T^\az$ and therefore $\kappa_\ell$, $\Omega_\varphi$, and
$\Phi_\ell$. For the Kerr-Newman spacetimes, it is natural to take
$T^\az$ as the stationary time Killing vector normalized to have length
$-1$ at infinity. Then for this choice of $T^\az$ and also gauging $\cA_\az$
so that $\Phi=0$ at infinity,
\bea
\kappa_\ell &=& \frac{R_\Delta^4 -  (Q_\Delta^4 + 4 J_\Delta^2)}{2 R_\Delta^3
\sqrt{(R_\Delta^2 +  Q_\Delta^2)^2 + 4 J_\Delta^2}} \\
\Omega_\varphi &=& \frac{ 2  J_\Delta }{R_\Delta 
\sqrt{(R_\Delta^2 +  Q_\Delta^2)^2 + 4 J_\Delta^2}} \\
\Phi_\ell &=& \frac{Q_\Delta}{R_\Delta} \frac{R_\Delta^2 +  Q_\Delta^2}{
\sqrt{(R_\Delta^2 +  Q_\Delta^2)^2 + 4 J_\Delta^2}}
\eea
where $R_\Delta$ is defined by $A_\Delta = 4 \pi R_\Delta^2$. It is 
easy to check that these take the forms required by (\ref{kdef}). Therefore
by the Smarr formula
\be
H_\Delta = \frac{\sqrt{(R_\Delta^2 + Q_\Delta^2)^2 + 4 J_\Delta^2}}{ 
2  R_\Delta}. \label{energy}
\ee
which of course is equal to the ADM energy in Kerr-Newman space (or the
Komar mass evaluated at infinity). 
Because they meet the conditions (\ref{kdef}) one is perfectly free to 
extend these definitions for 
$\kappa_\ell$, $\Omega_\varphi$, and $\Phi_\ell$ across all of phase
space and so effectively calibrate the invariants so that they will
match their natural values in the Kerr-Newman section of the phase space.

This calibration was first made in \cite{ABL} from a slightly different
perspective which it is illuminating to consider. There, no postulate 
was made about the form of $H_\Delta$, but instead it was simply required
that $\kappa_\ell$, $\Omega_\varphi$, and $\Phi_\ell$ take the above values
across the phase space so that they would match their natural definitions in 
the Kerr-Newman sector. Then, the first law was integrated and found to 
give rise to equation (\ref{energy}). Thus, from this perspective if one
calibrates $\kappa_\ell$, $\Omega_\varphi$, and $\Phi_\ell$ using the 
Kerr-Newman sector, then the given form of $H_\Delta$ follows from the
first law with no further assumptions. It is a derived quantity and though
the $\kappa_\ell$, $\Omega_\varphi$, and $\Phi_\ell$ have been specifically
chosen to match Kerr-Newman values, $H_\Delta$ has not but rather the 
formalism has forced it to take that form. 

\section{Comparison with Brown-York}
\label{BY}

The procedure used to calculate the variations in this paper is a
straightforward extension of the calculations found in the metric-based
Hamiltonian literature. In particular, it is very closely related
to both the author's PhD thesis \cite{mythesis} and the recent paper 
by Brown, York, and Lau \cite{BYLnonortho} 
which depart from earlier works by
calculating the variations directly from the three-surface based
Hamiltonian rather than from the four-dimensional action and so
obtain Hamiltonian rather than Lagrangian equations of motion
(though the boundary terms remain the same).
Thus, it is of interest to compare the current work with those 
earlier ones in the Brown-York tradition that derive quasilocal
energies and thermodynamics for regions with timelike boundaries. 
First, we examine the quasilocal energies.

\subsection{Quasilocal energy}
Let us compare the expressions for the quasilocal energies. On the
RRH, the NQLE can be written in integral form as
\be
H_\Delta = \int_{\Delta_t} d^2 x \ssg \left\{ 
\kappa_\ell/(4\pi) + \hat{V}^a \hat{\omega}_a/(4\pi) + E_\vdash 
(\Phi_\ell + 2 \hat{V}^\az \cA_\az)   
\right\}
\ee
while for a general $B_t$ in $B$, the corresponding
Hamiltonian boundary term may be written as
\be
H_B = \int_{B_t} d^2 x \ssg 
\left\{ \bN \bep  + \bV^\az \bj_\az + 
E_\vdash (2 \bN \Phi + 2 \hat{V}^\az \cA_\az) \right\},
\ee
where for now we have neglected the reference term arising from $I_0$.
Of course for $H_\Delta$, $T^\az$ is required to be tangent
to the null surface $\Delta$ while for $H_B$, $T^\az$ is required to be
tangent to the timelike surface $B$, so the timeflow vectors for the two
are not the same. That said, we can compare them by considering a surface
$\Delta_t = B_t$ formed when a null surface $\Delta$ and a timelike one
$B$ intersect. Then there is a different timeflow vector for each surface,
but nevertheless we now have the two QLEs defined on the same surface
and so can compare their functional forms. To faciliate this comparison, 
we assume that the each of the timeflow vectors projects to the same
shift vector $\hat{V}^\az$ in $\Delta_t = B_t$. . 

We consider the similarities, starting with the $\hat{V}^\az$
dependent terms. First note that
the electromagnetic angular momentum terms are identical. The
gravitational angular momentum terms are also very closely related. 
A straightforward calculation shows that 
\be
\hat{\omega}_\az = -\sigma_\az^\bz \ell^\cz \nabla_\bz \tl_\cz
= \sigma_\az^\bz \bu^\cz \nabla_\bz \bn_\cz + d_\az \xi = 8 \pi  
\bj_\az + d_\az \log \xi.
\ee
Thus, for cases where $\mathcal{L}_{\hat{V}} \xi = 0$, the angular momentum
terms are identical. Of course, given the near identical derivations
of the two QLEs this correspondence isn't too surprising, especially when
one considers that these are the components of the Hamiltonians 
``perpendicular'' to the boost that relates the two defining sets
of observers. See \cite{naked, mythesis} for the equivalent situation
discussed for sets of timelike observers that are boosted relative to each 
other. 

Less similar are the ``boost-dependent'' terms which we might expect to 
be different. Starting again with the electromagnetic term, we see that 
the two are functionally equivalent (up to the lapse dependence and factor of
two) though their Coulomb potentials are defined with respect to different 
normals. Most 
different though are the remaining two geometric terms which may be considered
the key terms in the expressions since they are all that remains
when the angular momentum and electromagnetic terms vanish. $H_\Delta$
depends on the acceleration of the null congruence of generators of 
$\Delta$ while $H_B$ depends on extrinsic curvature of the $B_t$ in 
$\Sigma_t$. The two are not equivalent, for the analogue of the extrinsic
curvature of $B_t$ is the expansion $\theta$ of $\Delta_t$, which is
set to zero by the defining conditions for non-expanding horizons. Further,
the corresponding acceleration $n^\az a_\az$ on $B$ appears only in 
the stress tensor $s^{ab}$ which in turn only shows up in calculating
the rate of change of the CQLE Hamiltonian if $B_t$ is not static
\cite{tidal,mythesis}.

Not surprisingly given the difference in the definitions of the two, 
their numerical values differ as well. For simplicity in the calculation I'll
drop the electromagnetic field (and the associated complications of
gauge choice) as well as the angular momentum (since it contributes 
in the same way to each term) and for this comparison work in
the Schwarzschild spacetime. Then, based on the calibration of the previous 
chapter
\be
H_\Delta = m,
\ee
where $m$ is the mass parameter for the black holes. Unfortunately, it is 
not quite so straightforward to give a value for $H_B$ as there are a 
variety of surfaces $B$, reference terms $\bep^0$, and lapses $\bN$ that
may be chosen. We will choose the so-called canonical QLE for our comparison
by setting $\bN = 1$ (and so normalizing $T^\az$ to length $-1$) and further
choose spherical $B_t$ surfaces so that we may use the standard embedding 
reference terms \cite{BY1}. Finally, we will choose to calculate the CQLE for 
a constant radius surface $B$ outside of $\Delta$ and then consider the limit
as $B \rightarrow \Delta$ (we could equally well consider a timelike surface 
$B$ intersecting $\Delta$ at $\Delta_t$ and then take the limit
as $B$ becomes null. The result is the same and dealt with in some
detail in \cite{nopaper,naked}). Then
\be
H^{CQLE}_B = r(1 - \sqrt{1-2m/r}),
\ee 
which goes to $m$ in the limit $r \rightarrow \infty$ and then monotonically
increases to $2m$ as $r \rightarrow 2m$ (though of course $B$ goes null
at $2m$ and so $H_B$ isn't properly defined right on the horizon). In the 
literature, this is usually taken to indicate that $-m$ units of gravitational
potential energy live between the horizon of the black hole and infinity.
From this point of view then, the fact that 
the  NQLE measures a mass of $m$ at the horizon could be taken as
indicating that it not only includes the external electromagnetic
``hair'' but also includes the corresponding gravitational ``hair''.

There is also a difference in the reference terms needed to normalize each
of the QLEs. Namely the reference term on $B$ is allowed to be any 
functional of $\gamma_{\az \bz}$ and $\gamma_\az^\bz \cA_\bz$ while 
$H_\Delta$ as we have seen is severely restricted in the form that 
it is allowed to take.
This difference is perhaps not so surprising when we examine the 
variational calculations more closely. While the mechanics of the calculation
are more-or-less identical, there are some significant differences in the 
boundary terms. Recall that in order for the variational principal to be
well defined for a timelike boundary, the full (induced) metric on that 
boundary must be fixed and unvarying. By contrast, for a null boundary, one 
need only say that it is a non-expanding horizon. The size of
cross-sections $\Delta_t$ on any non-expanding horizon is fixed in time for a
given horizon, but variations are allowed that will move to different 
horizons with different areas, charges, and angular momenta. Because there
are fewer fixed characterizing terms for an isolated horizon, there is 
correspondingly less freedom to add constants onto the action (and therefore
the Hamiltonian), to the extent that dual requirements that there be  a 
consistent Hamiltonian evolution and that $H_\Delta$ vanish when 
$A_\Delta, J_\Delta, \mbox{and } Q_\Delta \rightarrow 0$ essentially 
fixes what boundary terms can be added to the Hamiltonian in the null case.
By contrast, with the entire history and future of the boundary evolution 
defined 
in the time-like case, there is more freedom to add boundary terms. As long
as they are functionals of the boundary metric $\gamma_{\az \bz}$ they do 
not affect the Hamiltonian evolution. Thus, in this case there is an 
indeterminacy in the definition of the QLE that is not found for the NQLE on 
isolated horizons and one is free to choose ``reference'' terms appropriate
to a given situation (for example the embedding reference terms of \cite{BY1}
versus the intrinsic counterterms of \cite{intrinsic}).

It is of interest too, to compare the thermodynamics of the two formalisms. 
To do this however, we first need to note that the Brown-York paper \cite{BY2}
makes extensive use of the path-integral formalism of gravity to put
forward a theory of {\em thermodynamics} including derivations of 
the entropy and temperature. By contrast, in this paper we have
really been looking at isolated horizon {\em mechanics}. The difference
is that while we have tacitly assumed the surface-gravity/temperature 
and surface-area/entropy identities, they cannot be proved within
the classical framework used 
(but see \cite{IHentropy} for a canonical quantum gravity
calculation of the entropy of an isolated horizon). Thus, the aims of the
two are slightly different. Nevertheless with the variational
calculations underlying the two essentially the same it is of interest
to make the comparison, and so we now do that.

\subsection{Thermodynamics}
Again we start with the similarities between the isolated horizon and 
Brown-York approaches to thermodynamics. For the reader who is not familiar
with the Brown-York approach it is briefly
reviewed in appendix \ref{A}. Apart from the fact that they are both 
Hamiltonian based the other way in which they depart from textbook 
black hole thermodynamics is that they seek to study finite regions of
spacetime rather than entire spacetimes. While in the isolated horizon
case that region is contained by the isolated horizon itself, 
in the Brown-York case, the region contained by a timelike boundary some
distance from the horizon. Usually that boundary is taken to have
a static intrinsic metric (though that is not required by the formalism
as it is in the null case) and it is assumed that the fields inside of $B$ 
are stationary (though the fields outside of $B$ suffer no such restriction).

However, though they are closely related in these generalities, there are 
major differences. As has been noted and is discussed in appendix \ref{A}, 
Brown-York thermodynamics is
formulated with path-integral quantum gravity using the Euclidean 
instanton approximation. This instanton is constructed in the usual way 
by analytically continuing time to imaginary values and then periodically 
identifying it with a carefully chosen period. As a result of this 
process, the black hole horizon is reduced to a conical singularity in 
a complex manifold, and then further reduced to a smooth point no 
different from any other once an appropriate period is chosen. Thus, in a
certain sense, the region being studied is a finite part of a
``complex Euclidean'' manifold rather than a region of Lorentzian 
space containing a black hole horizon.

That said, the entropy of the region being studied corresponds to the 
action of the instanton (which approximates the full density of states
function). Its evaluation depends crucially in a balance between the
acceleration of $\bu^\az$ (as contained in the stress tensor $\bs^{ab}$ on 
the horizon)
and the chosen period of the imaginary time coordinate. This balance is fixed 
to eliminate conical singularities, but as a side effect gives the black hole
entropy as $A/4$. 

The thermodynamics is then extracted by considering variations of the
action that correspond to shifts through the phase space of black holes. 
The first law takes the form
\be
\delta (A/4)  = \int_{B_t} d^2 x \left\{
\beta \delta [ \ssg \bep] - \beta \Omega \delta [ \ssg \varphi^\az \bj_\az] 
+ \beta (\ssg s^{ab}/2 ) \delta \sigma_{ab} \right\},
\ee
derived as equation (\ref{firstlaw2}) in the appendix. Here the 
work terms are defined on the boundary
$B$ rather than the horizon $\Delta$. The inverse temperatures $\beta$
come from the lapse and shift integrated over the periodic time. The first
law takes an integral form in this case because the
zeroth law doesn't hold, which is to say that $\beta$ and $\Omega$ are not 
in general constant over $B$ and so cannot be pulled out of the integral.
Further, in addition to the work terms
that we found for RRHs, there is an additional work
term of the form $s^{ab} \delta \sigma_{ab}$ (the stress tensor 
contracted with area element distortions). 

\section{Discussion}
\label{discuss}

Hamiltonian investigations of general relativity have long been popular and in 
particular the recent analyses of isolated horizons are based on just such
an approach. Though up to now the isolated horizon work has been spinor or
tetrad based, the current paper has taken a metric-based approach. This will
be of interest for those who feel more comfortable with metrics rather than
connections and has also allowed a comparison of the results with the 
extant work on metric-based Hamiltonians, quasilocal energy, and quasilocal
thermodynamics. Further the calculations of this paper have been based
on direct variations of the Hamiltonian which in turn was found by a Legendre
transform of the action. This approach then serves to 
provide an alternative viewpoint from the symplectic geometry presentations 
of the existing isolated horizon papers,
though, as has been pointed out, the two are computationally equivalent.

We have seen that the isolated horizons fit very nicely into the metric-based
formalism with the required quantities that we want to fix naturally
occuring as boundary terms in the variational calculations. In particular
the expansion of the horizon and the electromagnetic matter flow across
it are naturally fixed, and these of course are key characteristics
of non-expanding horizons. In agreement with the earlier tetrad and spinor
approaches, we saw that in order for the variations of the action to vanish
across the phase space of RRH-containing-solutions to the equations of 
motion, the time evolution vector field $T^\az$ 
must be functionally dependent
on $A_\Delta$, $J_\Delta$, and $Q_\Delta$ which in turn means that the
secondary quantities $\kappa_\ell$, $\Omega_\varphi$, and $\Phi_\ell$ must
also be dependent on these invariants. As noted, this is a departure from 
ADM-type Hamiltonian treatments of general relativity where the time-flow 
vector field is an independent background structure and does not appear
in the four-dimensional covariant form of the action.

We proposed an $H_\Delta$ boundary 
term defined as a surface integral on $\Delta_t$ which gives rise to 
a Smarr-type formula for rigidly rotating horizons. Together with the
first law of RRH mechanics (which arises from the Hamiltonian formalism
whether or not we postulate that $H_\Delta$ takes this form), 
this formula then places a fairly strong 
restriction on the allowed functional forms of $\kappa_\ell$, 
$\Omega_\varphi$, and $\Phi_\ell$. The standard $\kappa_\ell$,
$\Omega_\varphi$, $\Phi_\ell$, and energy in Kerr-Newman spacetime 
satisfy these functional forms, and so may be used to calibrate these
quantities across phase space. It is important to remember however, that even 
without this calibration the Smarr formula and laws of isolated horizon 
mechanics continue to hold.

It is perhaps interesting to note that throughout this work, the assumption
that $\varphi^\az$ was a Killing vector of the RRH was only very weakly
used. In fact, the only purpose of the rotational symmetry assumption was to 
select out a vector field that would not be affected by the variations, and
so could be commuted with $\delta$ in the derivation leading up to 
the first law (equation (\ref{firstlaw})). If an alternate method could be
found for assigning vector fields in $\Delta_t$ across the phase space,
then one could easily reformulate all of this in terms of weakly isolated
horizons rather than RRHs.

The angular momentum arising from the Hamiltonian has previously been shown
to be closely linked to the equivalent Komar angular momentum in \cite{ABL}.
Here, we have also shown that our proposed Hamiltonian energy is similarly 
related to the Komar energy integral. In both cases, the purely gravitational
parts of the derived quantities exactly match the corresponding Komar
integral, while the electromagnetic parts do not. In turns out though that
for stationary spacetimes, those electromagnetic parts serve to 
make the quasilocal quantities defined on the horizon equal to the 
ADM/Bondi energy defined at infinity. Thus, in some sense these terms
include the electromagnetic (and gravitational depending on your point 
of view) ``hair'' of the hole in the quasilocal energy/angular 
momentum measured at the surface. 

This result is in contrast to the Brown-York quasilocal energy which is
based on essentially the same calculations as the current one, though with
timelike instead of null boundaries (in fact the non-orthogonal
version of their calculations has been been reproduced on our outer boundary 
$B$). For their CQLE the energy contained by a surface just outside the horizon
of a Schwarzschild hole is equal to twice the mass of the hole, 
while for the NQLE defined here it is equal to the mass itself. 
The CQLE, of course, may also be  defined off
the surface, in which case we find that it monotonically decreases to 
$m$ at infinity. This is often interpreted to mean that is it measuring
$-m$ units of gravitational potential energy outside the horizon. Taking
that point of view, one can then reconcile the two claims if one thinks
of the NQLE as including the ``gravitational hair'' just as it also
includes the electromagnetic. Then the CQLE could be thought of as 
measuring the bare mass not including the attendant gravitational 
fields outside the inner boundary. 

Again given the similar mathematical machinery underlying the two approaches
it is natural to examine the thermodynamics/mechanics to which they 
give rise. Though both the Brown-York and isolated horizon analyses are 
quasilocal in nature and Hamiltonian based, it was seen that in almost
all other respects they are quite different. In the first place of course,
Brown-York aim to provide a theory of {\em thermodynamics} while the isolated
horizon work is less ambitious and just seeks laws that are analogous to the
laws of thermodynamics. No attempt is made to prove the entropy/area and 
surface gravity/temperature associations. Second, the energy and angular 
momentum in the Brown-York approach are evaluated on a timelike surface
apart from the horizon while the corresponding quantities for RRHs are 
calculated directly on $\Delta$. Third while the zeroth law holds for
RRHs it does not for the timelike boundaries except in the static,
spherically symmetric case (see \cite{BY2}). Fourth, the actual 
use of the variational calculations in the two formalisms is 
quite different. The first law of RRHs essentially arises from a direct
variation of the Hamiltonian (identified with the energy). By contrast,
the Brown-York approach uses the variational principle and path-integral
gravity in the Euclidean approximation to calculate the total entropy
of the region of spacetime as the action of an appropriate instanton. 
Variation of that action the gives rise to the first law. 


It would be interesting to investigate more closely what 
restrictions the enforcement of strict Hamiltonian evolution might have 
on the allowed reference terms of the Brown-York action. Up to now they have 
only been considered the Lagrangian formulation. As has been noted however, 
the situation for the outer boundary terms is not the same as for the
horizon terms. In particular, on the inner boundary the general-covariance
of the action is broken by the inclusion of a time-flow dependent boundary 
term and as a result the time flow is restricted by the Hamiltonian formalism.
This is not the case at the outer boundary where the general covariance
has been preserved.
Nevertheless, it might
be worth thinking on this issue more closely to see if any restrictions
could be placed on the allowed reference terms.

\section{Acknowledgements}
This work was supported by the Natural Sciences and Engineering Research 
Council of Canada (NSERC). I would like to thank Chris Beetle, 
Steve Fairhurst, and Robert Mann for discussions regarding various parts of 
this work. 

\appendix
\section{Brown-York thermodynamics}
\label{A}
In this appendix we summarize the Brown-York approach to thermodynamics
(for more details see \cite{BY2}. For simplicity we 
will ignore the electromagnetic field. The quasilocal
region $M$ is taken to be same as that which we have considered in this
paper. That is, a timelike outer surface, a null inner surface, and
initial and final surfaces $\Sigma_1$ and $\Sigma_2$ that are spacelike
everywhere except right at their intersection with $\Delta$. 
However, the foliation $\Sigma_t$ (and therefore $\Sigma_1$ and $\Sigma_2$ 
as well) is taken to consist of stationary 
time slices and the time flow vector $T^\az$ is taken to be the stationary
Killing vector and an element of $T \Delta$ on that surface. Then we must 
have $N=0$ on $\Delta$ so that the surface will be null and further that 
surface will be non-expanding because $T^\az$ is the stationary Killing vector.
Further one assumes that the coordinate system is corotating with the horizon, 
so $V^\az/N$ vanishes there. 
Meanwhile, at the outer boundary, $T^\az$ is assumed to lie in $TB$, so that
boundary is stationary as well. Further on $B$, microcanonical boundary
conditions are imposed. That is $\bep$, $\bj^\az$, and $\sigma_{\az \bz}$ 
are fixed and therefore the quasilocal energy, angular momentum, and area
of the outer boundary are fixed. Thus, looking at the system as a whole, 
it is isolated from the outside with fixed energy, angular momentum, and 
surface area and contains a non-expanding horizon. 

By analogy with non-gravitational physics, the path integral formulation of 
quantum gravity then says that we can find the density of states of the
system by evaluating the path integral
\be
{\it v}(\bep,\bj ,\sigma) = \int \mathcal{D} H_P e^{i I_E / \hbar},
\ee
where the integral is over all possible spacetime configurations $H_P$ 
(not just solutions to the equations of motion) that are periodic in 
coordinate time period $P = t_2 - t_1$ and which have a fixed data
$\bep$, $\bj_\az$, and $\sigma_{\az \bz}$ on the outer boundary. 
Because of these microcanonical boundary conditions, the entropy
of the system is approximately equal to $\log \it{v}(\bep, \bj, \sigma)$. 

The catch, of course, is that no one knows how to evaluate such 
gravitational path integrals exactly. Instead, the steepest descent
approximation is used. That is, it is assumed that the entire integral
considered above may be approximated by
\be
{\it v}(\bep,\bj ,\sigma) = e^{i I^\star / \hbar},
\ee
where $I^\star$ is the microcanonical action of a complex and periodic 
solution to the equations of motion that also satisfies the required 
boundary conditions. Such a solution may be easily constructed by
analytically continuing 
$N \rightarrow - i N$ and $V^\az \rightarrow - i V^\az$ 
for the original solution that we are studying. Now after this rotation,
$\Delta$ is no longer null. In fact because $V^\az/N \rightarrow 0$ there
it actually becomes Euclidean. Further, because $N=0$ the complex spacetime
closes up at $\Delta$ and a conical singularity appears unless the time period
$P$ is carefully chosen. Specifically, we must have 
\be
\label{constraint}
\frac{2 \pi}{P} = \bn^\az \nabla_\az N = \lim_{r \rightarrow r_\Delta}
4 \pi N (\sigma_{\az \bz} s^{\az \bz}).   
\ee
Thus, the microcanonical action should be one that fixes the lapse, shift,
and pressure $s^{\az \bz} \sigma_{\az \bz}$ on the inner horizon and
$\bep$, $\bj_\az$, and $\sigma_{\az \bz}$ on the outer. From the considerations
of section \ref{calcs} it is easy to see that the appropriate action is
\bea
I_{mc} &=& \left\{ \mbox{ momentum and constraint terms } \right\} \\
    && + \int_{t1}^{t2} dt \int_{B_t} d^2 x \ssg
 \left\{ \bN (\bep - \bep^0) - \hat{V}^a \bj_a - 
         \frac{N}{2} s^{ab} \sigma_{ab} \right\}. \nn            
\eea
Then, for a stationary complex solution to the Einstein equations that
is constrained so that $N=0$, $V^\az/N = 0$, and 
equation (\ref{constraint}) holds, this is equal to $-i A_\Delta/4$.
Thus the standard entropy result is obtained.

The first law is then derived by considering $\delta I_{mc}$. Again
leaning on the results from section \ref{calcs} and now assuming a 
rotational symmetry, we have
\be
\label{firstlaw2}
\delta (A/4)  = \int_{B_t} d^2 x \left\{
\beta \delta [ \ssg \bep] - \beta \Omega \delta [ \ssg \varphi^\az \bj_\az] 
+ \beta (\ssg s^{ab}/2 ) \delta \sigma_{ab} \right\},
\ee
where we've altered the notation slightly to match that used in this paper.
$\beta = P$ is the inverse temperature and it comes out of the time-integrated 
lapse and shift terms.

\end{document}